\documentclass[a4paper,fleqn]{cas-dc}

\usepackage[authoryear]{natbib}
\usepackage{amssymb}
\usepackage{graphicx}
\usepackage{epstopdf}
\usepackage{graphicx}
\usepackage{algorithm}
\usepackage{subfigure}
\usepackage[noend]{algpseudocode}
\usepackage{amsthm}
\usepackage{url}
\usepackage{enumerate}
\usepackage{amsmath}
\usepackage{multirow}
\usepackage{float}	
\usepackage{flushend}
\usepackage{ragged2e}
\usepackage{color}
\usepackage{booktabs} 
\usepackage{ulem}

\usepackage{xurl}

\def\tsc#1{\csdef{#1}{\textsc{\lowercase{#1}}\xspace}}
\tsc{WGM}
\tsc{QE}

\begin{document}
\let\WriteBookmarks\relax
\def\floatpagepagefraction{1}
\def\textpagefraction{.001}
\shorttitle{Neural-FEBI: Accurate Function Identification in Ethereum Virtual Machine Bytecode}
\shortauthors{J. He et~al.}

\title [mode = title]{Neural-FEBI: Accurate Function Identification in Ethereum Virtual Machine Bytecode $^{**}$}                      
%
%
%
\author[1]{Jiahao He}[orcid=0000-0002-6837-2682]
\fnmark[1]
\ead{jiahaohe@webank.com}

\author[2]{Shuangyin Li}[orcid=0000-0001-6404-3438]
\fnmark[1]\cormark[1]
\ead{shuangyinli@scnu.edu.cn}

\author[3]{Xinming Wang}[orcid=0000-0003-0416-9471]
\ead{wangxinming@lakala.com}

\author[4]{Shing-Chi Cheung}[orcid=0000-0002-3508-7172]
\ead{scc@cse.ust.hk}

\author[2]{Gansen Zhao}
\cormark[1]
\ead{gzhao@m.scnu.edu.cn}

\author[2]{Jinji Yang}
\ead{yangjj@scnu.edu.cn}

\address[1]{WeBank Co., Ltd, Block A Building 7 Shenzhenwan Keji Eco Park, Shenzhen, China}

\address[2]{School of Computer Science, South China Normal University, Zhongshan Road No. 55, Guangzhou, China}

\address[3]{Lakala Co. Ltd., Building D1, Zhongguancun Yihao, Beiqing Road, Beijing, China}

\address[4]{Department of Computer Science and Engineering, The Hong Kong University of Science and Technology, Clear Water Bay, Kowloon, Hong Kong}

\fntext[1]{These authors contributed equally to this work and should be considered co-first authors.}
\cortext[1]{Corresponding author }
\cortext[2]{© 2023. This manuscript version is made available under the CC-BY-NC-ND 4.0 license https://creativecommons.org/licenses/by-nc-nd/4.0/}

\begin{abstract} 
Millions of smart contracts have been deployed onto the Ethereum platform, posing potential attack subjects.
Therefore, analyzing contract binaries is vital since their sources are  unavailable, involving identification comprising function entry identification and detecting its boundaries. 
Such boundaries are critical to many smart contract applications, e.g. reverse engineering and profiling. 
Unfortunately, it is challenging to identify functions from these stripped contract binaries due to the lack of internal function call statements and the compiler-inducing instruction reshuffling. 
Recently, several existing works excessively relied on a set of handcrafted heuristic rules which impose several faults. 
To address this issue, we propose a novel neural network-based framework for EVM bytecode Function Entries and Boundaries Identification (neural-FEBI) that does not rely on a fixed set of handcrafted rules. 
Instead, it used a two-level bi-Long Short-Term Memory network and a Conditional Random Field network to locate the function entries.
The suggested framework also devises a control flow traversal algorithm to determine the code segments reachable from the function entry as its boundary.
Several experiments on 38,996 publicly available smart contracts collected as binary demonstrate that neural-FEBI confirms the lowest and highest F1-scores for the function entries identification task across different datasets of 88.3 to 99.7, respectively. 
Its performance on the function boundary identification task is also increased from 79.4\% to 97.1\% compared with state-of-the-art. 
We further demonstrate that the identified function information can be used to construct more accurate intra-procedural CFGs and call graphs.
The experimental results confirm that the proposed framework significantly outperforms state-of-the-art, often based on handcrafted heuristic rules. 
\end{abstract}



\maketitle

\section{Introduction}
\label{sec:introduction}
Ethereum smart contracts \citep{buterin2014next} are programs that automate blockchain transactions. The industry has increasingly used them for finance, insurance, identity management, and supply chain management. 
Unfortunately, smart contracts are historically error-prone and pose attack subjects because they often help handle and transfer valuable assets. 
Therefore, smart contracts have attracted the attention of security applications, including automated auditing and automated rewriting to enhance their security. 
However, many smart contracts do not have readily linkable public source code available. 
Even when source code is available, experience has shown that the semantics of the executed binary code can differ from the source code \citep{BalakrishnanR10}. 
Instead, security applications had to focus on introducing binary analysis to understand the intent of contract binaries.

The primary challenge of such binary analysis for smart contracts is the lack of high-level semantic structure within the binaries. 
This is because the compilers discard the high-level semantic structure during the compilation process. 
The function is the basic and key structural piece.
Therefore, function identification is a preliminary step in many binary analysis techniques.
The information on the function boundaries provides the basis for recovering the high-level static constructs such as the call graphs and intra-procedural control flow graphs (CFGs) \citep{grech2019gigahorse, brent2018vandal}. These constructs are critical to many applications, such as context and path profiling, widely used to detect intrusion for vulnerable contracts \citep{wang2019contractguard}.
In addition, such functionality provides concert targets for instrumentation or modification. 
Two common instrumentation operations are instrumenting the entries of all basic blocks of a given function and instrumenting function entries and exits. 
Once the function boundaries are inaccurate, client tools may instrument in the wrong places, causing users to miss attacks or report false alarms.

Nevertheless, binary-level contracts are often represented as a sequence of hexadecimal numbers, namely Ethereum Virtual Machine (EVM) bytecode \citep{wood2014ethereum}, without any information about how parts are grouped into functions. 
Identifying functions for this stripped EVM bytecode requires considering the unique characteristics of the EVM bytecode, which are not found in other conventional bytecodes. 
The internal functions are not top-priority in the EVM as all contract functions are embedded in a single stream of instructions as low-level jumps for internal calls and returns.
This makes it challenging to differentiate their internal call statements from other control statements.
Consequently, the function entries are not identifiable by simple destination analyses of the internal call-sites.
Moreover, the Solidity compiler utilizes \textit{Deduplication} optimization to reduce the contracts’ code size by sharing the duplicated code segments.
This induces overlapping functions, including functions sharing code and non-contiguous functions, typically confusing the existing tools in labeling the accurate function boundaries.

Recent studies \citep{grech2019gigahorse,brent2018vandal,zhou2018erays,Elipmoc} 
have focused on identifying functions for the EVM bytecode by using handcrafted heuristic rules. 
For example, they suggest a few rules for identifying call-sites and determining the function entry points. 
However, these trivial rules are insufficient for complex constructs induced by the compiler optimization process and the frequent changes in different implementations of the compiler. 
Therefore, they fail to address the challenges mentioned above.
For example, in our experiments, the function boundaries identified by Gigahorse \citep{grech2019gigahorse}, a widely adopted decompiler, achieve an F1 score of about 53.1\% - 79.4\% across different datasets.
Such failures may harm the client security applications since the client security applications built on top of these tools assume that the underlying tools provide reliable outputs.

Therefore, this paper proposes a novel EVM bytecode Function Entries and Boundaries Identification framework based on neural network (Neural-FEBI \footnote{The tool and its dataset are available on https://github.com/ouerum/ neural-FEBI)}. 
In this method, we treat the function entries identification as a 0/1 classification task, where each address in a contract is judged to be a function entry or not. 
The functions have many valuables and routine operations at the entry, e.g., initializing the local variables. 
The widely used \textit{modifier} feature supported by Solidity smart contracts is compiled into similar bytecode and is always embedded before or after the entry of functions. 
These clues can then be used to effectively distinguish the function entries effectively. 
We train an FSI-LSTM-CRF (Function Starts Identification based on bi-Long Short-Term Memory network and Conditional Random Field) neural network to utilize the bytes of the EVM bytecode as the input and predict whether each location is a function's entry. 
 With the function entries, the framework designs a control flow traversal algorithm to recognize the boundary of each function.
 
 The proposed algorithm uses a two-step approach to identify the internal call-sites by which the proposed framework can easily differentiate the inter- and intra-procedural control flows. 
 In the first step, we use public function entries reported in the dispatcher and the candidates we identified by the FSI-LSTM-CRF neural network to detect internal call-sites.
 If a jump target is known as the function entry, it is a candidate internal call-site. 
 In the last step, several strategies are designed to be applied during the control flow traversal aiming to filter the internal call-sites and identify the correct boundaries.

To evaluate our framework's performance, we conduct abundant experiments on a large-scale real-world Ethereum smart contract from EtherScan, where 1,189,567 source codes are collected.
The contracts are then deduped, and the ground truth is extracted by instrumenting the Solidity compiler. 
A subset with 38,996 unique contracts is then used to evaluate the performance of the function entries task and boundaries identification. 
Based on contracts across different compilers and optimization, the experimental results show that our framework achieves minimum and maximum values of 88.3\% and 99.7\% on the F1-score, for the task of function entries identification. 
For the task of function boundaries identification, the proposed neural-FEBI also achieves high performances on the F1-score in the range of 84.6\% to 97.1\%. 
Our proposed framework also outperforms the methods using handcrafted fixed rules, such as Elipmoc, Gigahorse, and Vandal.
We further demonstrate the accuracy of the call graph and intra-procedural control flow graph (CFG) construction with function boundaries identified by neural-FEBI. 

In summary, this paper makes the following major contributions:
\begin{enumerate}
	\item We devise an automated technique to effectively identify function entries in EVM bytecode using a novel FSI-LSTM-CRF neural network.
	\item We propose a two-step algorithm based on control flow traversal to locate all associated code segments as the function boundaries.
	\item We implement our method as an automated tool. Compared with the available tools on real-world contracts, the proposed significantly improves on several evaluation metrics, such as F1-score, precision, and recall. 
	\item We further extend the neural-FEBI to construct the intra-procedural CFGs and call graphs with the aid of identified function boundaries.
	\item We release the related datasets to the public, providing the first publicly available dataset for function entries and boundaries identification in the EVM bytecode.
\end{enumerate}

The rest of this paper is organized as follows. Section~\ref{background} introduces the background on smart contracts and function identification. Section~\ref{secchal} gives details of challenges in function identification in EVM bytecode. Section~\ref{secEVMFBD} presents the design of neural-FEBI. After reporting the experimental results in Section~\ref{experiments}, we close with the related works in Section~\ref{relatedworks} and the conclusion in Section~\ref{conclusion}.

\section{Background}\label{background}
\subsection{EVM and Smart Contracts}

EVM is the core environment for executing the Ethereum smart contracts. 
The EVM adopts the stack machine architectures which uses a 256 bits stack to record the operands and computation results of the EVM instructions. 
The EVM also provides persistent global storage and memory for temporary data. 
Moreover, the EVM bytecode is the only form of storage for smart contracts in the Ethereum blockchain. 
All contracts written in a high-level language must be compiled into the EVM bytecode, a sequence of hexadecimal numbers. 
Each byte of the EVM bytecode corresponds to an EVM instruction, and the only exception is the PUSH instruction, pushes a constant value into the operand stack. 
The encoded constant arguments carried by PUSH instructions are directly placed after the PUSH. 
Hence, the sequence of hexadecimal code can be translated into a sequence of EVM instructions which is more familiar for client analysis.

Ethereum smart contracts are mostly written in the Solidity programming language. 
At the source level, contracts written in Solidity contain declarations of state variables, definitions of functions, constructors, and modifiers. 
Solidity contracts are compiled into EVM bytecode for deployment by the Solidity compiler. 
A piece of code called dispatcher is then added as the entry of the entire contract. 
Once other Ethereum accounts invoke a contract, the contract starts executing at the dispatcher, decodes the message and jumps into an appropriate function.
A Modifier is a piece of bytecode used to modify the behavior of a function and embedded as a wrapper for the function body.

\begin{figure}[t]
	\center
	\subfigure[unoptimized]{
		\label{unoptimized_cfg}
		\includegraphics[width=2.5in]{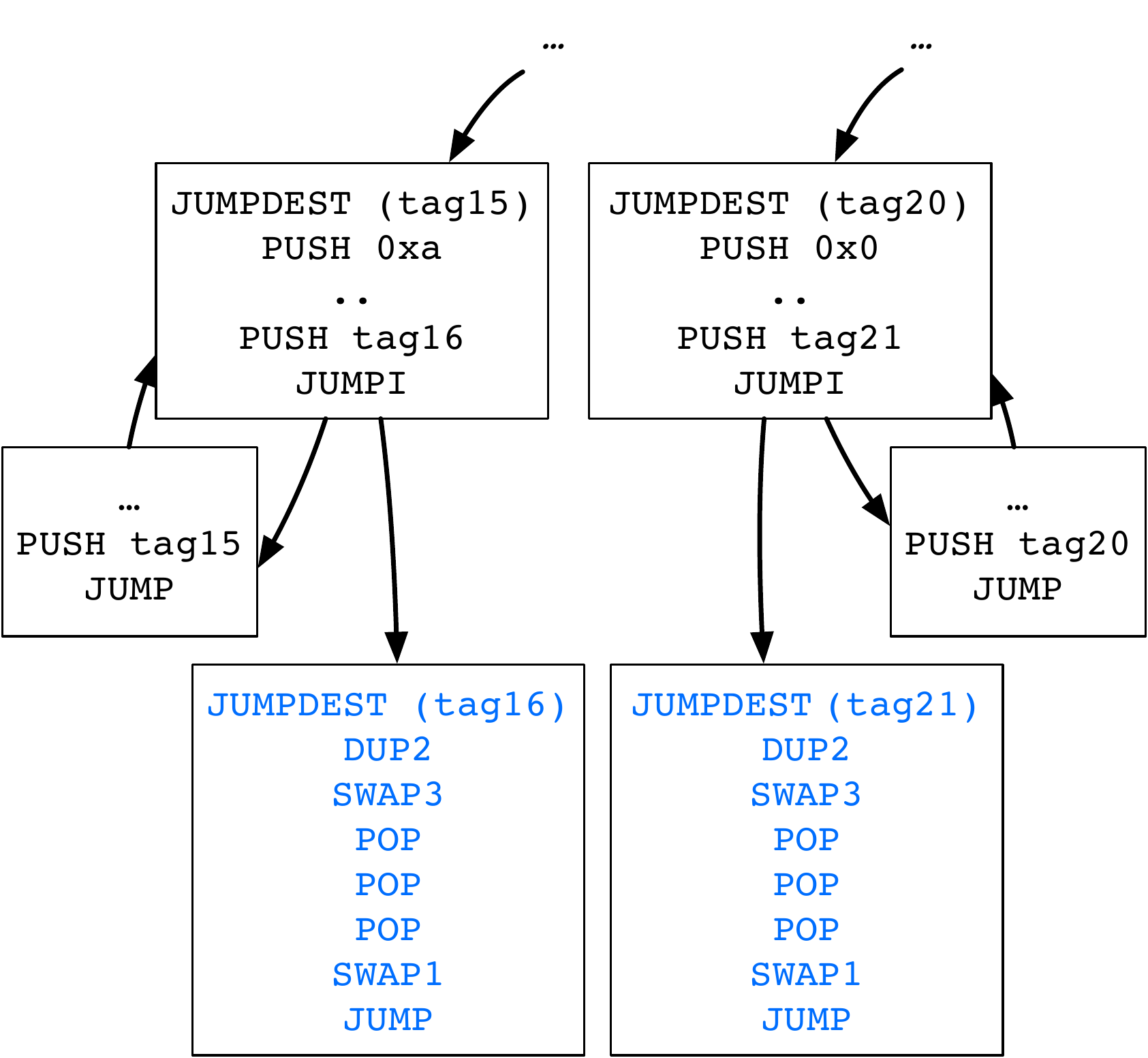}
	}
	\subfigure[optimized]{
		\label{optimized_cfg}
		\includegraphics[width=2.5in]{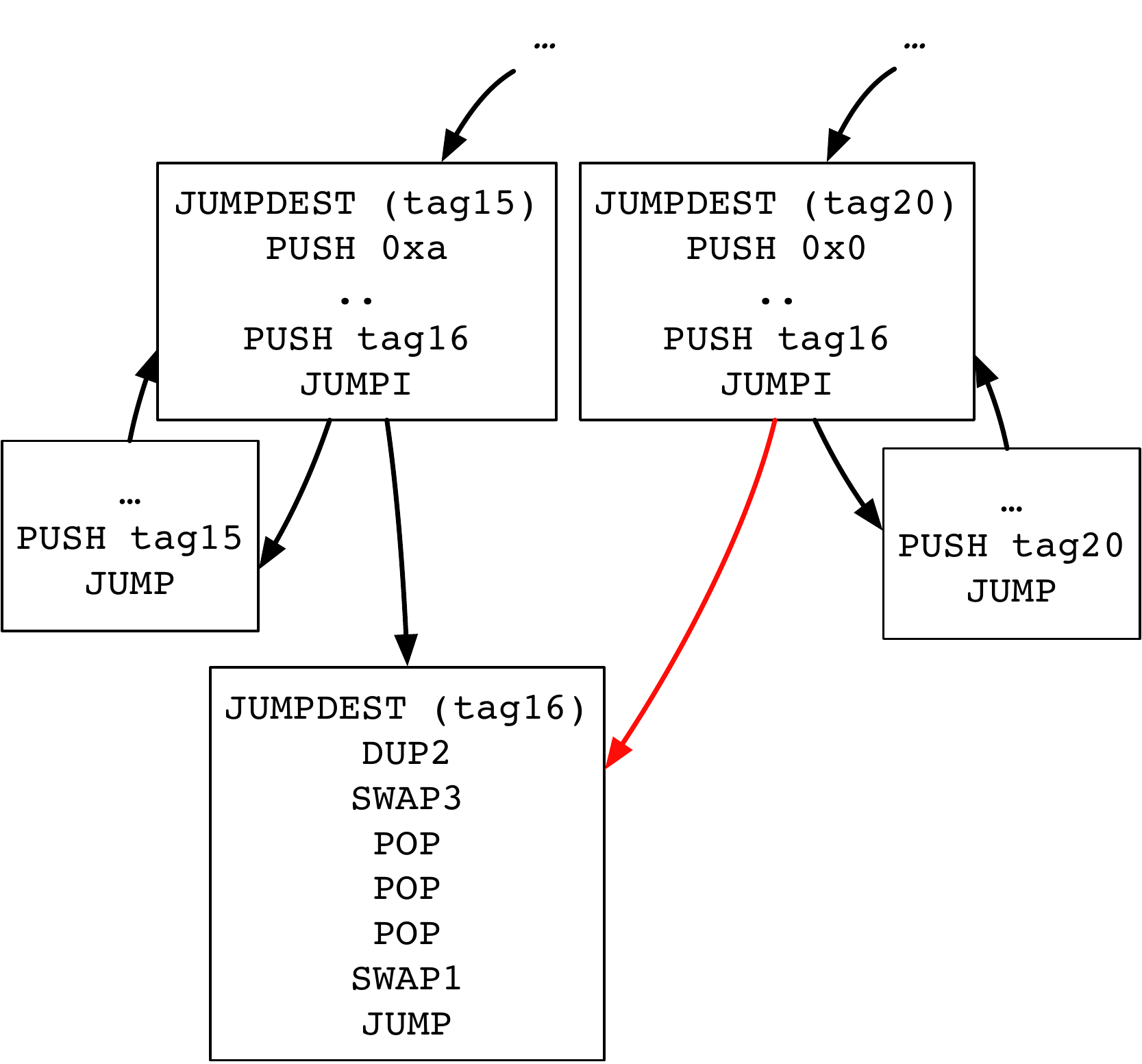}
	}
	\caption{Change induced by the block deduplication on the CFG.}
	\label{optimization_main}
\end{figure}

Ethereum smart contract execution must be replicated among nodes in the network. 
To prevent resource exhaustion, users must pay gas fees for each instruction every time they deploy or invoke a smart contract. 
The Solidity compiler introduces \textit{JUMPDEST Remover}, \textit{Peephole}, \textit{Deduplication}, \textit{CSE} to minimize the cost of smart contract deployment and execution. 
A major optimizer adopted by the Solidity compiler is \textit{Deduplication} (see Figure. \ref{unoptimized_cfg}).
If the target blocks of two jump instructions have a duplicated code, the Deduplication optimizer chooses the target block of one jump instruction and then redirects the other jump instruction to the chosen block. 
As illustrated in Figure. \ref{unoptimized_cfg}, \textit{tag16} and \textit{tag21} refer to two pieces of the duplicated block. 
The optimizer chooses \textit{tag16} and redirects the jump of \textit{tag21} to \textit{tag16} by modifying the PUSH \textit{tag21} and PUSH \textit{tag16}. 
Block deduplication is common in the EVM bytecode. 
In practice, the bytecode of more than 97\% of real-life contracts has been optimized by \textit{Deduplication}.

\subsection{Functions Identification}
Function identification aims to determine a set of functions that exist in the stripped EVM bytecode, which comprises the superset of function entries and boundaries identification. 
If the debug information of the symbolic tag exists, determining the function becomes trivial. 
However, analyzing the stripped bytecode is more challenging when not assisted by debugging information.

The function entries identification task is the basis of determining function boundaries. 
Several previous works proposed various solutions, from simple heuristic rules to approaches using neural networks, with
many focusing on manually designed instruction patterns for conventional bytecode, which were then exploited to identify the entries of functions. 
In addition, many sophisticated tools based on this technique have been widely used in reverse engineering, including IDA Pro \footnote{https://www.hex-rays.com/products/ida/}, Jakstab \footnote{http://www.jakstab.org/} and OllyDbg \footnote{http://www.ollydbg.de/}. 
For the EVM bytecode, \citep{brent2018vandal, grech2019gigahorse, zhou2018erays} have designed different heuristic rules to identify the function entries in the EVM bytecodes. 
However, \citep{rosenblum2008learning} and \citep{bao2014byteweight} demonstrated that these approaches are insufficient as they cannot adapt to the variations in the various compiling and optimization approaches. 

Some recent works also used supervised learning techniques and neural networks to automatically learn the features for identifying the function entries. \citep{rosenblum2008learning, bao2014byteweight, shin2015recognizing}
For instance, \citep{bao2014byteweight} proposed BYTEWEIGHT, a machine-learning-based approach for the x86 function entries identification task. 
\citep{shin2015recognizing} also inferred the recurrent neural network \citep{elman1990finding} could identify function entries in binaries with higher accuracy and efficiency.

For the function boundaries identification task, some tools represented the boundaries as a tuple of entry and end offset. 
However, this oversimplified definition of the boundaries is not suitable for the complex function structure in the stripped bytecode{,} conventional bytecode and EVM bytecode. 
Therefore, the functions must be represented as a set of bytes. 
The function boundaries need to recover recursively added instructions that are reachable from these function entries. 
The only vital challenge of this task is to differentiate an inter-procedural and intra-procedural control flow. 
This task is trivial in the conventional bytecode as their call-sites are easy to be identified correctly. 
Recently, BYTEWEIGHT\citep{bao2014byteweight}, BAP\citep{brumley2011bap}, and Dyninst\citep{meng2016binary} have achieved desirable performances in x86 bytecode for this task.
  
Although such tools have obtained acceptable results for other conventional bytecode, they are not suitable for the same task in the EVM bytecodes.
This work focuses on function identification in the stripped EVM bytecode, a sequence of hexadecimal numbers. 
Here the assumption is that the bytecodes of a smart contract are accessible, but no information is available about function ${f_1, ..., f_n}$.
We separate the task into two sub-tasks. First, we identify the function entries, and secondly obtain all bytes that belong to this function.

\newdefinition{rmk}{Definition}
\begin{rmk}
	Given a smart contract in a binary form, we find a set of $\{s_1,s_2,..,s_n\}$, where each $s_i$ is the first byte offset from $0x00$ in each function $f_i$.
\end{rmk}

\begin{rmk}
	The function boundaries identification problem is to output a set of $\{F_1, F_2,..., F_n\}$ where each $F_i$ is a set of byte offset corresponding to the body of a high-level function $f_i$.
\end{rmk}

The above problem definition forms the difficulty of the task, where the identification of function entries is easier than the identification of the full function boundaries.

\subsection{The basics of Neural Network}
This section presents the basis of the neural network used in our proposed neural-FEBI. 

Long Short-Term Memory (LSTM) is a recurrent neural network (RNN) architecture to address the vanishing and exploding gradient problems of conventional RNNs. 
The LSTM contains special units called memory blocks in the recurrent hidden layer, which contain cells with self-connections storing the temporal state of the networks in addition to special multiplicative units called \textit{gates} to control the information flow. 
More precisely, \textit{input} and \textit{output gates} controlled the input and output flow.
A \textit{forgot gate} was added to the memory block to scale the internal state of the cell before adding it as input to the cell through the self-recurrent connection of the cell, therefore adaptively forgetting the cell's memory.

A bidirectional LSTM (bi-LSTM) \citep{graves2013speech}  is a sequence-processing model that comprises two LSTMs: one taking the input in a forward direction and the other in a backward direction. 
As such, the bi-LSTM can efficiently use past features (via forward states) and future features (via backward states) for a specific timeframe.
In the proposed neural-FEBI, two levels of bi-LSTM are utilized to capture more information and identify EVM bytecode function entries.
The instruction-level bi-LSTM operates on individual instruction disassembled from bytecode, while the block-level one operates on the basic blocks grouped by the instructions.

The LSTM produces a position-by-position distribution over output labels and thus can suffer from the same label bias problem that motivates the Conditional Random Fields (CRFs). 
In contrast to the local normalized models, the CRF is a sequence model comprising a single exponential model for the joint probability of the entire sequence of labels given the observation sequence. 
In other words, CRF can predict labels based on not just one position but also the neighborhood.
Commonly, Viterbi decoding is used to find the most optimal tag sequence from the scores computed by a CRF.

\section{Challenges}\label{secchal}
As discussed above, there are three main challenges in identifying functions for the EVM bytecode. 
These challenges often confuse the existing tools, making it difficult for analysts to understand the intentions of smart contracts. 
In this section, we examine the details of such challenges.

\subsection*{Challenge \uppercase\expandafter{\romannumeral1}: Missing symbol}
To minimize the deployment cost, the EVM bytecode comprises a sequence of hexadecimal numbers optimized to contain only the information necessary for execution ("stripped code"). Therefore, the EVM bytecode does not provide the symbolic tagging of the function entries. 
Without this information, the function identification task for the EVM bytecode becomes significantly challenging.

\subsection*{Challenge \uppercase\expandafter{\romannumeral2}: Implicit Function Calls}
There are no special instructions to distinguish internal (intra-contract) function calls and returns in the EVM instruction set. 
In other words, to call an internal function, the EVM pushes the return address onto the operand stack, pushes arguments, pushes the called function entry address, and performs a jump into the called function. 
To return, the EVM performs an indirect jump using the address already pushed into the operand stack. 
Therefore, the functions declared in the source code are fused in the instructions sequence. 
If the analysis cannot identify the call-sites amongst many instructions, the control flow will be wrongly labeled as intra-procedural, and the code segment in the called function will be mistakenly labeled as part of the caller.

Observing implicit function calls may suggest a pattern-match approach to identifying call-sites and determining the target of such calls as function entry points. 
The algorithm proposed in \citep{grech2019gigahorse, brent2018vandal} is rather simplistic for the task of function identification: a call-site must push a return address and arguments in the same basic block, which jumps to the entry of a called function. 
These approaches work for simple cases, such as the one presented below, which call the function \textit{foo} with argument \textit{0xff} and \textit{0x0}.
\begin{figure}[h]
	\centering
	\includegraphics[width=3.2in]{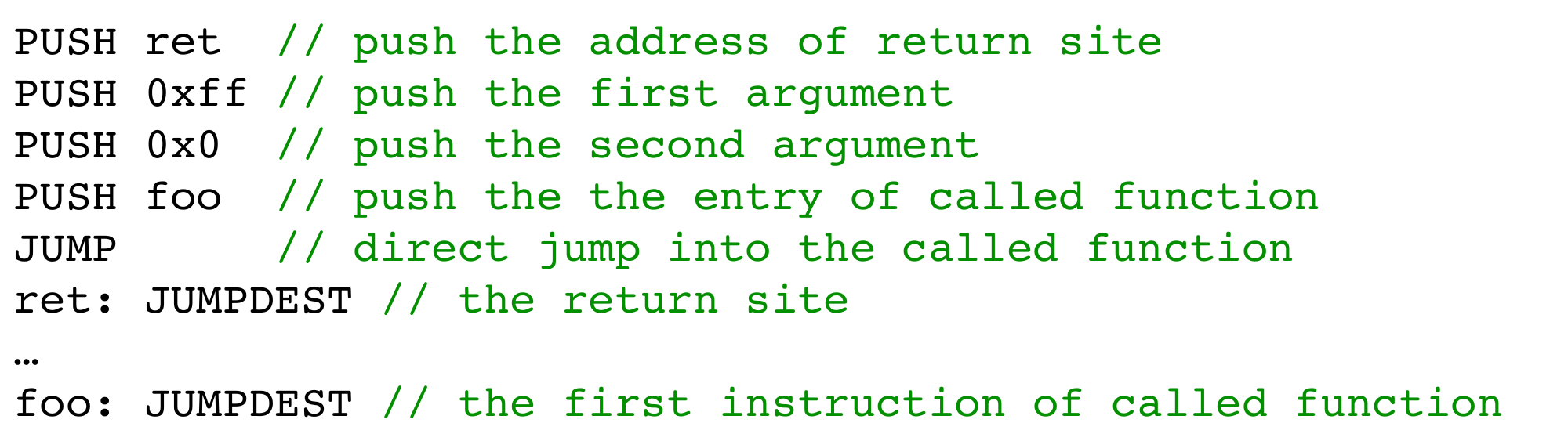}
	\caption{The instructions of function calls \textit{foo(0xff,0x0)}.}
	\label{simple function calls}
\end{figure}

However, these approaches are insufficient as they cannot adapt to the complex arguments in the internal function calls: they can not detect the return address pushed in long before the call-site basic block.
To motivate this case, we consider the following example in EVM bytecode that implements the high-level calls \textit{foo(a>b?a:b)}, whose argument is a triple expression. 
This high-level call is compiled into three different basic blocks, and the call-site basic block only provides the feature of jumping to the entry of the called function.
Therefore, these simple rules are insufficient as they cannot adapt to complex arguments in the function calls.

\begin{figure}[h]
	\centering
	\includegraphics[width=3.2in]{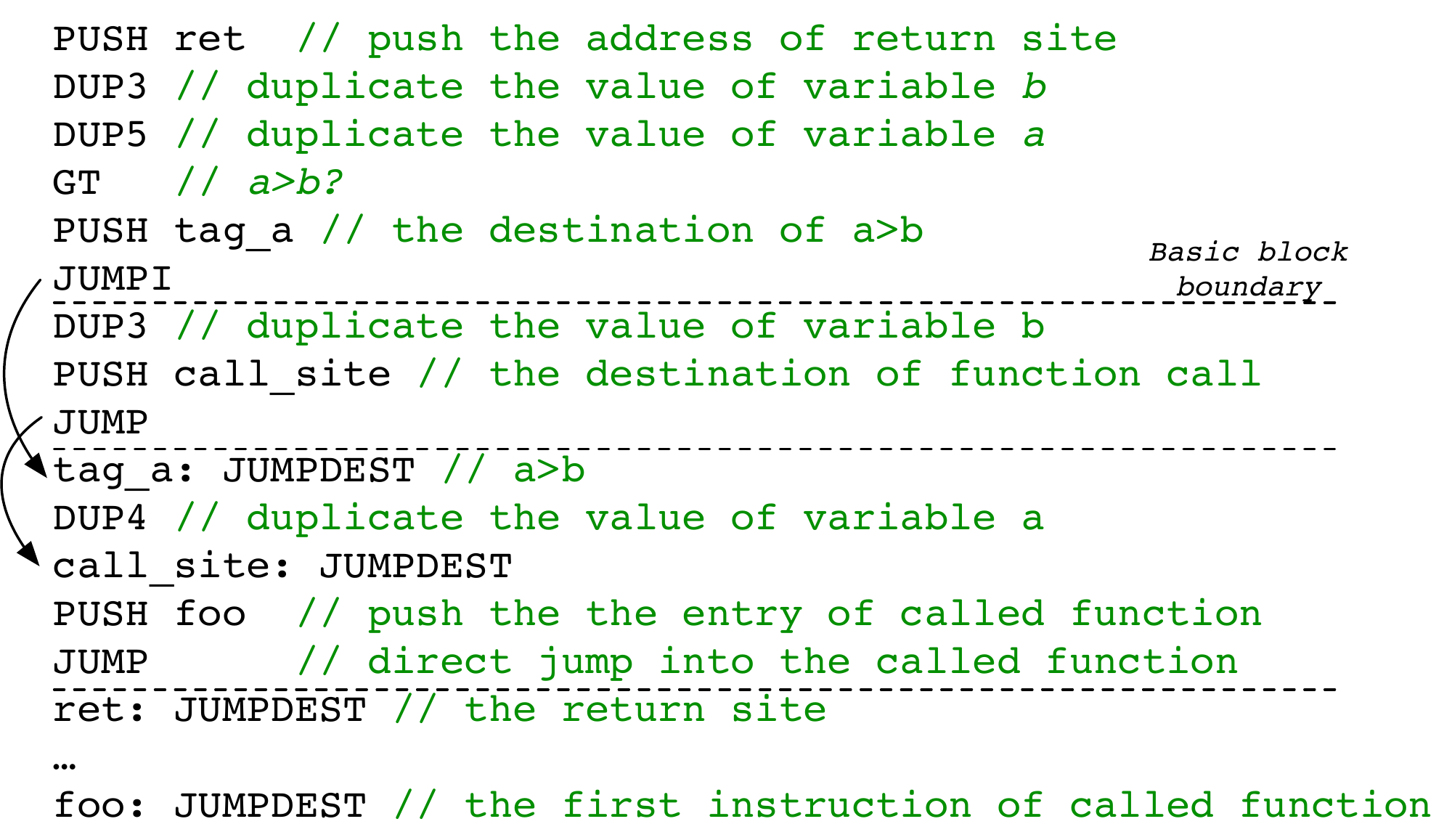}
	\caption{The instructions of function calls \textit{foo(a>b?a:b)}.}
	\label{complex function calls}
\end{figure}

\subsection*{Challenge \uppercase\expandafter{\romannumeral3}: Instruction Reshuffle}
\begin{figure}[h]
	\centering
	\includegraphics[width=2.5in]{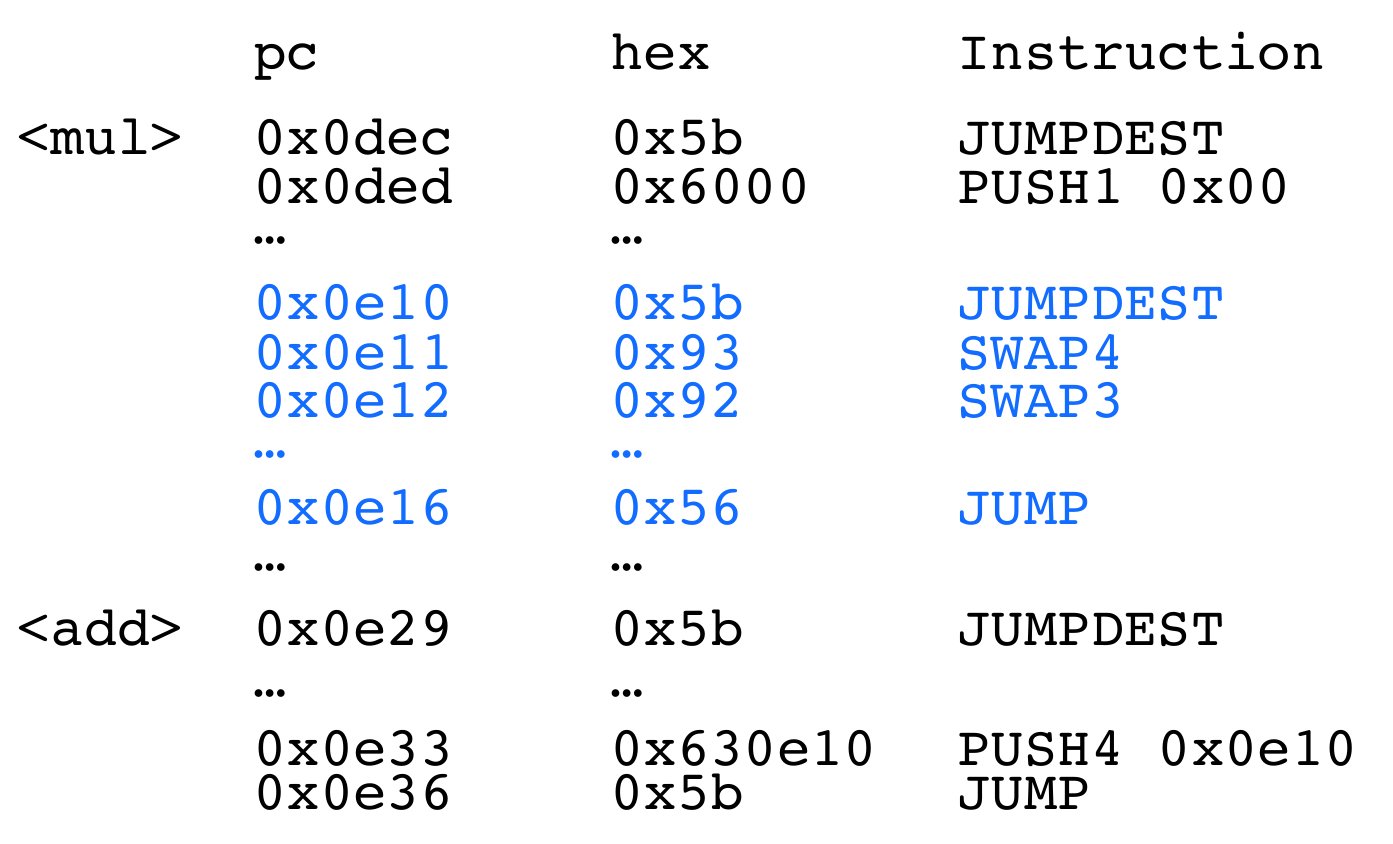}
	\caption{Two function sharing code: an example from a real-life contract, SiaCashCoin.}
	\label{complex function}
\end{figure}

Functions written by Solidity may have some common functionalities which lead to the same instruction sequence after compilation. 
\textit{Deduplication} optimization is then used to share similar code segments amongst the functions, further inducing non-contiguous function codes. 
Figure. \ref{complex function} illustrates an example, where two functions share code where \textit{mul} and \textit{add} both use the code from \textit{0x0e10} to \textit{0x0e10}. 
Note that \textit{add} is also a non-contiguous function for which part of the code lays before the entry point. 
A common mistake in the existing tools is to assume that a code segment can only belong to one function \citep{grech2019gigahorse}.

\begin{figure*}[t]
	\centering
	\includegraphics[width=6.4in]{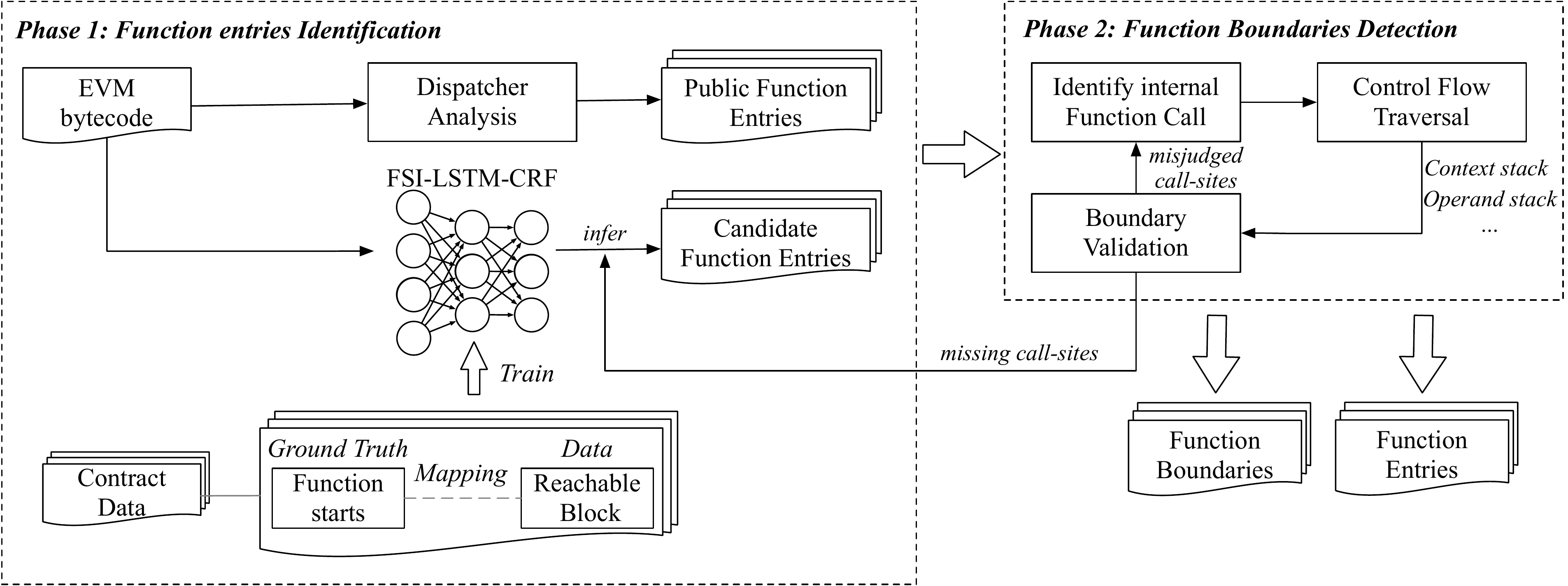}
	\caption{Overview: the workflow of neural-FEBI framework.}
	\label{workflow}
\end{figure*}

\section{The Proposed EVM bytecode Function Entries and Boundaries Identification Framework}\label{secEVMFBD}

This section proposes a novel  EVM bytecode Function Entries and Boundaries Identification framework based on neural network (neural-FEBI). 
The proposed method addresses the challenges presented in Section \ref{secchal}. 

Figure. \ref{workflow} illustrates a schematic overview of the proposed framework. 
In phase~1, we split the task of function entry identification into two subproblems, including public and internal entry identification. 
Given the EVM bytecode, the public function entries are directly identified by analyzing the contracts dispatcher and do not require further confirmation. 
Trained on sufficient cases, a neural network named FSI-LSTM-CRF is then established to predict the probability of being an internal function entry for each address in the contract. 
The addresses are collected as the candidate function entries which are then confirmed according to their probability. 

Once the entries are prepared, we recognize the reachable instructions as the body of the function via intra-procedural control transfer and by performing control flow traversal. 
To differentiate the internal call-sites and simple control statements, we assume that the instructions with control flow transfer to a candidate (or a public) entry are identified as the internal call-sites. 
During the control flow traversal, the information of the context and operand stacks are used to validate the identified boundaries.
If the information proves that the boundaries contain mistakes, the boundaries of these functions are reidentified by avoiding the misjudged call-sites or lowering the threshold of the candidate entries in phase~1.
This enables discovering the underlying function entries and missing call-sites and confirms the candidate entries predicted since the functions which are never called will be removed.

\subsection{Function Entries Identification}
In the proposed neural-FEBI, the task of function entry identification is to obtain the location of the first byte of each function. 
To minimize the deployment cost, the symbols corresponding to the function in EVM bytecode are removed, which is the basis for determining the function entries. 
Our handling of the function entries identification task is divided into two subproblems, the public and the internal function entries identification. 
Note that the dispatching patterns are clear. Therefore, the method of public function identification is rather straightforward. 
For internal function entries, we train a neural network model named FSI-LSTM-CRF to extract the features of common pattern sequences before or after the function entries. 
This is to infer the probability that an address is a function entry. 
This approach is also adaptable to variations in compilers and optimization without requiring handcrafted rules.

\subsubsection{Public Function Entries Identification}
\begin{figure}[h]
	\centering
	\includegraphics[width=3.2in]{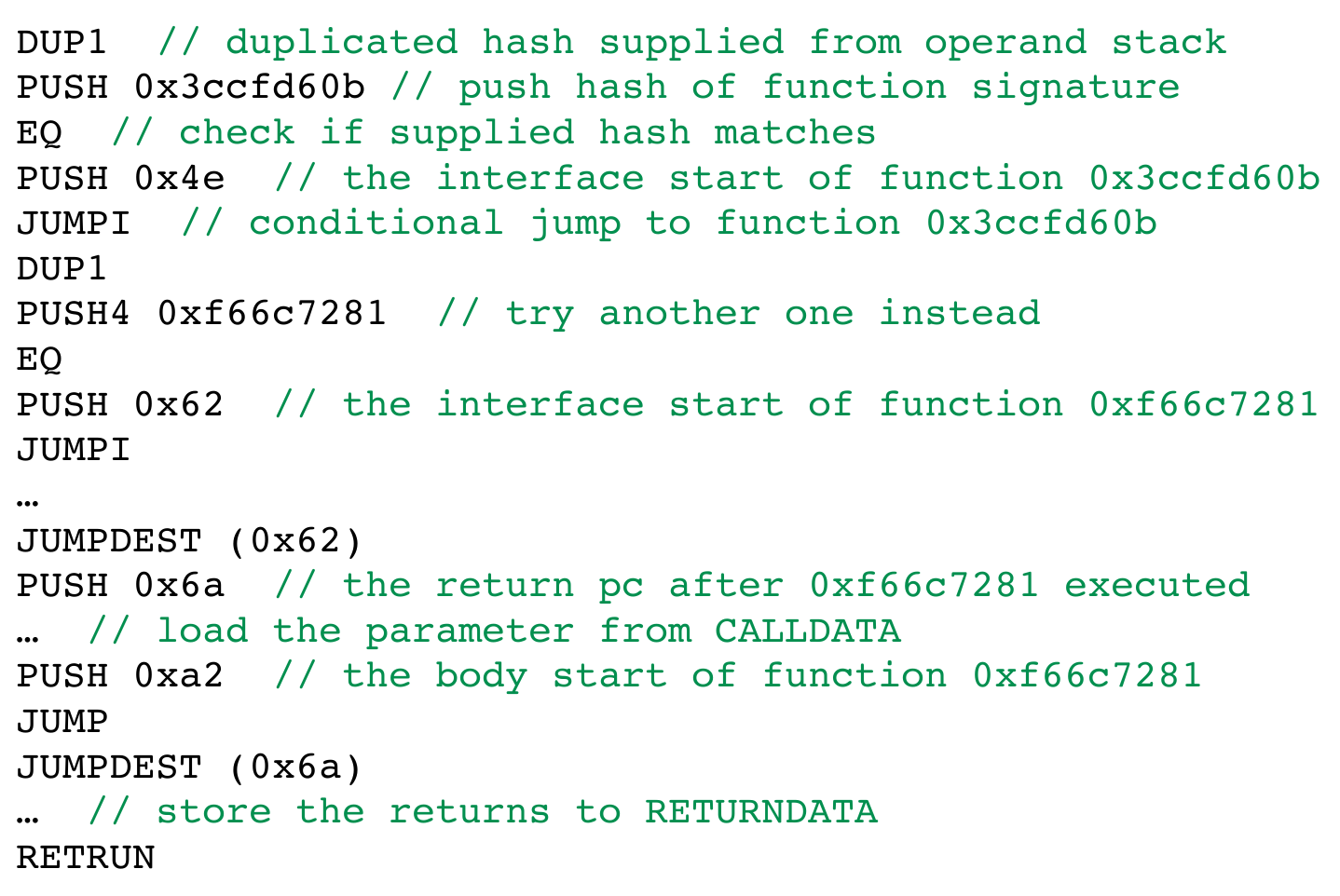}
	\caption{Dispatching pattern in the dispatcher for public function entries.}
	\label{public function}
\end{figure}

Each Ethereum contract is accompanied by an Application Binary Interface (ABI) description. 
The ABI enlists the contract’s public functions and their parameters in JSON format. 
The ABI description enables us to readily identify all public function entries by analyzing the dispatcher at the entrance of overall smart contracts. 
Figure. \ref{public function} provides an example of the code segment at the dispatcher with comments. 
The example contains two public functions. 
The first function has a function descriptor of \textit{0x3ccfd60b} with an entry at \textit{0x4e} and the second function has a function descriptor of \textit{0xf66c7281} with an entry at \textit{0x62}.

\begin{figure*}[t]
	\centering
	\includegraphics[width=6.5in]{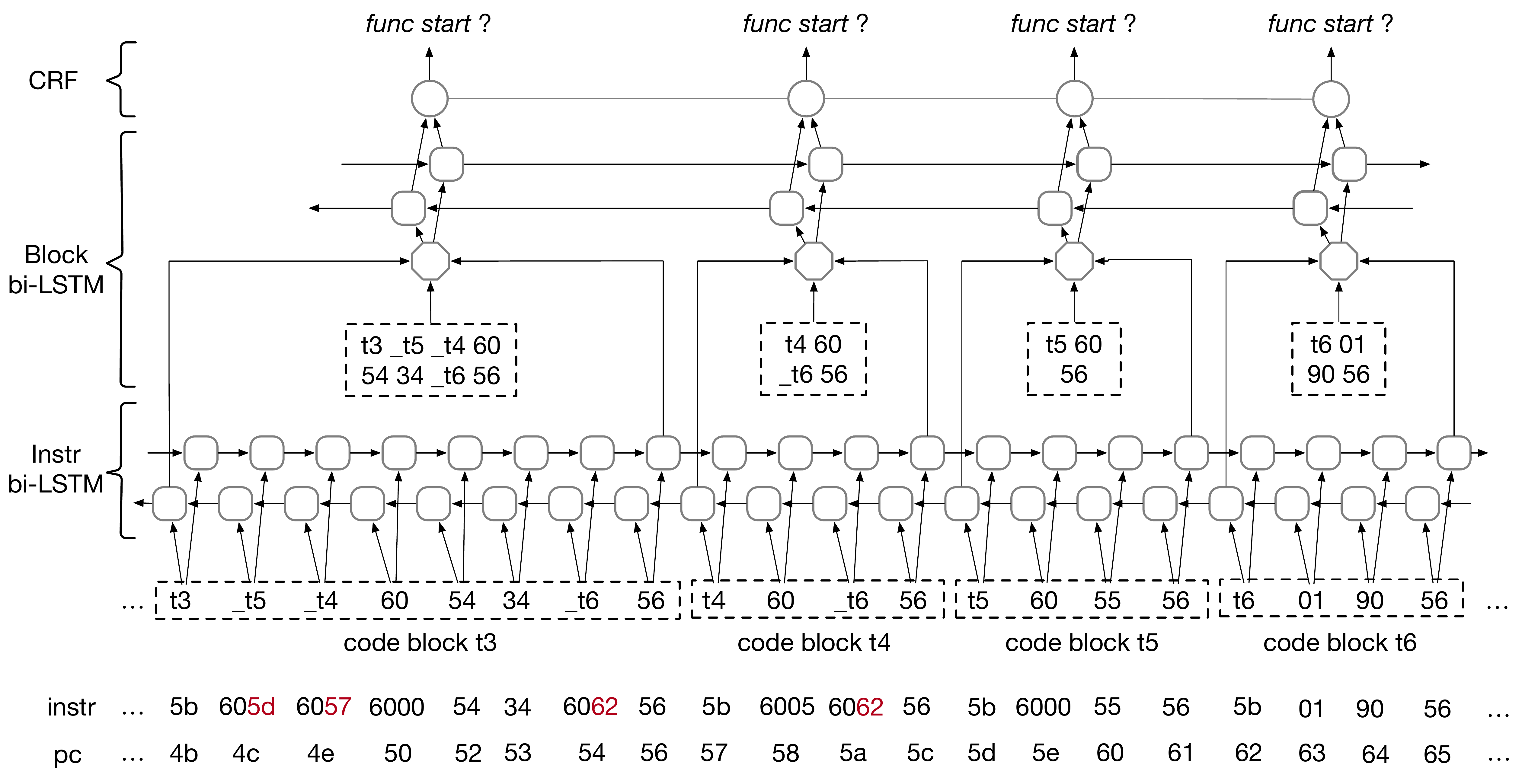}
	\caption{The architecture for internal function entry identification in the proposed neural-FEBI.}
	\label{model architecture}
\end{figure*}

Note that the entries identified in the dispatcher are not real entries of the public functions. 
We denote these entries as the interface entries of the public function. 
Upon jumping into an interface entry, a piece of code exists to decode the parameters from the message and then jump into the body of the public function. 
Once the execution of the code in the body is complete, the contract returns to this piece of code and packs the return values. 
Once other functions call the public functions by an internal call, they directly enter the body of the public function instead of jumping to the corresponding interface entry. 
The entries of the body in public functions are denoted as body entries.
 As seen in Figure. \ref{public function}, for the public function the body entry of \textit{0x3ccfd60b} is \textit{0xa2}.

\subsubsection{Internal Function Entries Identification}
A key challenge in function entry identification concerns the internal functions whose entries are not kept by the EVM bytecode. 
Unlike other conventional bytecodes (e.g., JVM bytecode), EVM has no special instructions for internal calls. 
In other words, all the internal calls are translated to low-level jumps with constant values. 
This challenges the identification of the entries for the internal functions. 
Fortunately, we found that the functions have many valuables and routine operations at the entry, such as local variable initialization or setting up a stack frame. 
Moreover, the similar widely used \textit{modifer} (e.g. \textit{owner}, \textit{mutex}) to change the behavior of functions are always embedded before or after the entry of the function. 
These common operations result in common instruction sequences. 
We can find the function entry points if we can learn the programmed rules of these instructions.
Such rules can be learned using a neural network.

Here, we propose a neural network architecture pipeline that goes through each valid instruction to determine if they are the function entries. 
The proposed FSI-LSTM-CRF framework is illustrated in Figure. \ref{model architecture}, comprising an instruction-level bi-LSTM, a block-level bi-LSTM, and a conditional random field layer.

\textbf{Data Preprocessing:}~
The input required for this particular step is the raw bytecode instructions extracted from the training dataset.
Each instruction contains a hex of one opcode. 
The only exception is the \textit{PUSH} instructions, followed by an operand.
These bytecodes must be passed through the preprocessing stage to be ready before being fed to the model. 
Here are the following steps through which the bytecode instructions are preprocessed:

\textit{1) Removal of Operands:} The EVM instruction set contains only 138 opcodes, but we receive an uneven number of instructions after disassembling the real-world smart contracts because any operands can follow the \textit{PUSH} instructions.
As such, removing some operands from the \textit{PUSH} instruction is required to limit the data size input to the model. 
In neural-FEBI, we remove the operands of \textit{PUSH}, which is only used for arithmetic, and keep the operands used as destinations for jumping. 

\textit{2) Labeling of Code Blocks:} 
After the first step, the next thing required in the dataset creation process is partitioning the instructions into a reachable block and providing a class to each reachable block. 
The reachable blocks are derived through two simple rules. Instructions that halt the execution (i.e. \textit{STOP} or \textit{REVERT}) mark block exit, while the special instruction \textit{JUMPDEST} marks block entry. 
Note that the reachable blocks are different from basic blocks because the basic blocks require linear execution while the reachable blocks are not. 
In other words, the \textit{JUMP/JUMPI} instructions can appear in the middle of a \textit{reachable block}. 
In our scenario, the reachable blocks whose first byte are function entry points are labeled as 1 while the rest are defined as 0.
 
Using reachable block instead of the common basic block makes the task of function entry identification simple. 
In the EVM, the \textit{JUMPDEST} is an instruction with no operations that marks the instruction’s location as a valid jump target address. 
If the EVM jumps to a non-\textit{JUMPDEST} instruction, it throws an exception and rolls back the current transaction. 
As such, each function entry must be a \textit{JUMPDEST} instruction, so it can be called by jumping.
For this reason, the function entry point may appear in the first instruction of the reachable block. 
It is possible to work at the most common basic-block level. 
However, most basic blocks do not start with \textit{JUMPDEST} instruction and these basic blocks are impossible to function entries for EVM smart contracts.

\textbf{Instruction Level bi-LSTM:}~
The instruction level model is trained purely based on unannotated sequence data, and it is nevertheless able to capture the underlying style and structure of each reachable block. 
In the task of function entry identification, it is especially useful since the part of the instructions sequence in the reachable block can often yield important clues on whether it is a function entry or not. 
Instead of predicting the next instruction through instruction-level LSTM, we adjust the task to make the next prediction for the next reachable block at the boundaries of the reachable block and capture the potential features. 
This leverages parts of the instructions sequence information to predict the next reachable block. 

Accordingly, we use two LSTM units to capture this information in forward and backward information. 
As shown at the bottom of Figure. \ref{model architecture}, the instruction-level bi-LSTM is inputted with a sequence of instructions after removing operands, and it outputs the prediction of the next instruction when the boundaries of a reachable block are met.

\textbf{Block Level bi-LSTM:}~
As a block level structure, we adopt a bi-LSTM to capture information in both directions. 
This bi-LSTM encodes the feature of block-level and instruction-level into new features at each reachable block containing information about the block and its neighborhood.
We concatenate the embeddings of the reachable block and the outputs of instruction-level bi-LSTM (including two directions) and feed them to block-level bi-LSTM. 
Since the reachable block constructed by a sequence of instructions does not belong to any language vocabulary, the existing fine-tuned pre-trained word embedding \citep{pennington2014glove, mikolov2013distributed} will not be suitable for the task
Hence, we co-train the whole block-level layer.

\textbf{Conditional Random Field Layer:}~
Instead of using a single linear layer to transform the output of block-level bi-LSTM into the scores of function entry, here we use neighbor tag information to predict if the current reachable block is a function entry. 
For example, if a reachable block is a function entry, it is highly unlikely for the next reachable block to be a function entry. 
Therefore, we leverage a conditional random field (CRF) \citep{lafferty2001conditional} to label a sequence of reachable blocks in the contracts jointly. 
In our model, the CRF layer takes the outputs of block-level bi-LSTM as inputs and transfers them into emission and transition scores for each block.
The emission scores measure the likelihood that the block is a function entry, and the transition scores measure how likely it is to transition from one tag to another. 
We use the Viterbi Decoding to construct the most optimal tag sequence for function entries by using the aggregated value of the emission and transition scores from the CRF layer.
Each decoding step identifies reachable blocks with probabilities larger than a dynamic threshold as function entries. 
We employ the information provided in the next phase to adjust the threshold and filter some predicted entries in the decoding step.

\subsection{Function Boundaries Identification}
Once the function entries are available, we perform control flow traversal to find the code segments that are reachable from the entries in the proposed neural-FEBI. 
When an internal call-site is reached, we then ignore the execution of the called functions but continue to process to its return-site.
In a conventional programming language, the task {of function boundary} is not complex since we can easily differentiate the inter-procedural control flow transfer from the existing \textit{CALL} instruction for internal function calls. 
However, the oversimplified design of the EVM instruction set makes it difficult to distinguish the internal call-sites from other intra-procedural control statements. 
This problem is similar to the tail calls \citep{clinger1998proper}  in the x86 bytecode, where all internal calls become "tail calls" in the EVM bytecode.

\subsubsection{Approach Overview}
The algorithm we implemented in the proposed neural-FEBI comprises the following steps:
\begin{enumerate}
	\item \textit{Inferring Call/Return Instructions: }  Given an instruction sequence grouped in basic blocks, the potential internal call and return instructions are identified with the local analysis within the basic blocks.
	\item \textit{Reachable Instructions Identification: } The operand and context stacks are subject to symbolic execution to identify the reachable instructions from the function entries.
	\item \textit{Boundaries Validation: } When the return-sites are met in control flow traversal, the information provided by the operand stack and context stack is used to confirm the correctness of potential call-sites and predicted function entries.
\end{enumerate}

\subsubsection{Inferring Internal Call/Return Instructions}
A basic block is a maximal linear sequence of instructions with a single entry and exit.
The task of aggregating the instructions is derived from a well-known rule. 
Each basic block either starts with a \textit{JUMPDEST} or ends with an instruction that alters control flow (i.e. branches or halts).

Once the instruction sequence is divided into basic blocks, we perform a local analysis of the operand stack effects of basic blocks. 
This operation is to compute the jump destination of the \textit{JUMP/JUMPI} instructions by analyzing the instructions within basic blocks.
Since the jump destination is not always available at the top of the operand stack in local analysis, we can identify two types of jumps: \textit{direct} and \textit{indirect}.
A \textit{direct jump} is a \textit{JUMP/JUMPI} whose jump destination is pushed in the same basic block so that its target is easy to solve just by looking for the value in the preceding \textit{PUSH} instruction.
The destination of the \textit{indirect jump} is not immediate to compute with the analysis within the boundary of basic blocks as they have been pushed long before.
The potential internal call and return instructions can be inferred with local analysis. 
A \textit{direct jump} with the destination target to the predicted function entries is considered a potential internal call instruction.
Any \textit{indirect jump} is determined as the internal return instruction because the compiler uses them to return to the destinations already pushed in the call-site.

\subsubsection{Reachable instructions Identification}
Once the public function and the predicted entries are available, a control flow traversal algorithm is performed to associate each instruction reachable from there with these entries.
Starting with a function entry, the algorithm executes the operand stack and context stack symbolically by walking the partially built control flow graph (CFG). 
The analysis walks through the CFG using a DFS (Depth-First Search) by updating the state of these stacks.
To simplify the implementation, we only symbolically execute the instructions dealing with the jump destination, including \textit{PUSH}, \textit{AND}, \textit{DUP}, and \textit{SWAP}. 
These instructions either prepare the jump address on the top of the operand stack or compute the actual jump address with a mask code (i.e. \textit{0xffffffff}).
For other instructions, the algorithm only performs the \textit{pop} and \textit{push} operations for the operand stack with an unknown symbol $\top$. 

Another important aspect is that the identified call/return instructions are met in control flow traversal, resulting in the context stack updating.  
According to the mentioned challenges, we can not identify the return-site of each call instruction, as there are no special labels to distinguish the jump destination and other values in the operand stack.
In other words, we can not identify the corresponding return address for each internal function call.
As such, we have to track the execution of the called function and push the calling context into the context stack when the call instructions are reached.
Similarly, the element on the top of the context stack is popped when the return instructions are reached.
With the aid of the context stack, it is easy to identify the current instructions that belong to which function. 

The detailed algorithm for boundary identification is presented in Algorithm .\ref{alg::static analysis}. 
The algorithm takes the public function entries and the candidate entries predicted in the previous phase in the previous phase as input.
In line 3, the queue \textit{WorkList} used for DFS is initialized: it contains an entry $e$ of function, an empty context stack $ctx$, and an operand stack $stack_{op}$ pushed several special symbols $\$$. 
These special symbols are prepared as the parameters and a return address because the operand stack may use these data when it updates.
Then the algorithm proceeds with the execution of the symbolic stack until the \textit{WorkList} is empty.
In lines 6-7, the reached instructions are inlined in the function that starts from $e$ when the context stack $ctx$ is empty.
The algorithm symbolically executes the instruction and updates the state of the operand stack $stack_{op}$ at line 8.
The next instruction wait for processing is computed according to the following rules:
\begin{itemize}
	\item \textit{JUMP} makes an unconditional jump, which transfers control to the instruction at the target address held by the top element of the stack.
	\item \textit{JUMPI} makes a conditional jump. The instruction at the target address held by the top element of the stack and the fall-through instruction are waiting for exploration.
	\item \textit{STOP} and other halt instructions terminate program execution, and there are no successors.
	\item In any other case, the fall-through instruction is considered.
\end{itemize}
In lines 13-14, the context stack $ctx$ is updated by pushing the current calling context $(entry, pc)$ into it when the internal function calls are reached.
Similarly, the top element on $ctx$ is popped when the algorithm reaches the return statement in lines 9-10.
The algorithm adds the successors and the updated stacks to the $Worklist$ that waits for further exploration in line 15.

The proposed algorithm is path-sensitive.
This algorithm does not join the states with the same address in cases where the control flow includes different paths.
Path sensitivity, in general, is prohibitively expensive because of the path explosion. 
Several precautions must be taken to limit path explosion and ensure the termination of the proposed algorithm.
We merge the symbolic state if the $pc$ and the jump destination stack $stack_{tag}$ are identical.
The jump destination stack is a subset of operand stack $stack_{op}$, in which the jump destinations in the contents of $stack_{op}$ are extracted. 
Due to the challenge of determining jump destinations and other values in the operand stack, we have to collect the operands equal to the address of \textit{JUMPDEST} instructions as $stack_{tag}$ in our implementation.
Although the path explosion cannot be fully avoided, our experimental results show that only a small scale of overtime contracts exists across different datasets. 
As such, if the $(pc', stack_{tag})$ is not visited before, we stop exploring it further.

\renewcommand{\algorithmicrequire}{\textbf{Input:}}  
\renewcommand{\algorithmicensure}{\textbf{Output:}} 

\begin{algorithm}[h]
	\caption{The algorithm of the neural-FEBI for the function boundaries identification in EVM bytecode.} 
	\label{alg::static analysis}
	\begin{algorithmic}[1]
		\Require
		$entries$;
		$instructions$;
		\Ensure
		the function boundaries of each function.
		\For {$e$ in $entries$}
			\State prepare a $op\_stack$ with a few $\$$s
			\State $Worklist=[(s, [\,], stack_{op})]$
			\While {$len(W)$!=$0$}
				\State $pc$, $ctx$, $stack_{op} = Worklist.pop()$
				\If {$ctx$ is empty}
					\State record $pc$ as associated with $e$
				\EndIf
				\State $next$, $stack_{op}'$ = Analysis($pc$, $stack_{op}$)
				\If {$pc$ is indirect jump}
					\State $ctx.pop()$
				\EndIf
				\For {$pc'$ in $next$}
					\If {$(pc',stack_{op}')$ is unvisited}
						\If $pc'$ is call instruction
							\State add $(entry, pc)$ to $ctx$
						\EndIf
						\State add $(pc',ctx,stack_{op}')$ to $Worklist$
					\EndIf
				\EndFor
			\EndWhile
		\EndFor
	\end{algorithmic}
\end{algorithm}

\subsubsection{Boundary Validation}
Even after identifying of the function entries in the EVM bytecode, there is not enough information to differentiate the internal call-sites and the simple control statement. 
Furthermore, the predicted function entries may contain false positives and false negatives resulting in misjudgment and missing the call-sites. 
To address this issue, we propose two two-step approaches. 
In the first step, the proposed algorithm identifies call-sites if their targets are either in the external entries or the predicted candidate.
In the second step, the context stack $ctx$ and operand stack $stack_{op}$ are verified upon reaching a return instruction. 
This information provides evidence enabling the detection of incorrectly identified boundaries. 
The details of validation are explained below.

\begin{figure}[t]
	\centering
	\includegraphics[width=2.5in]{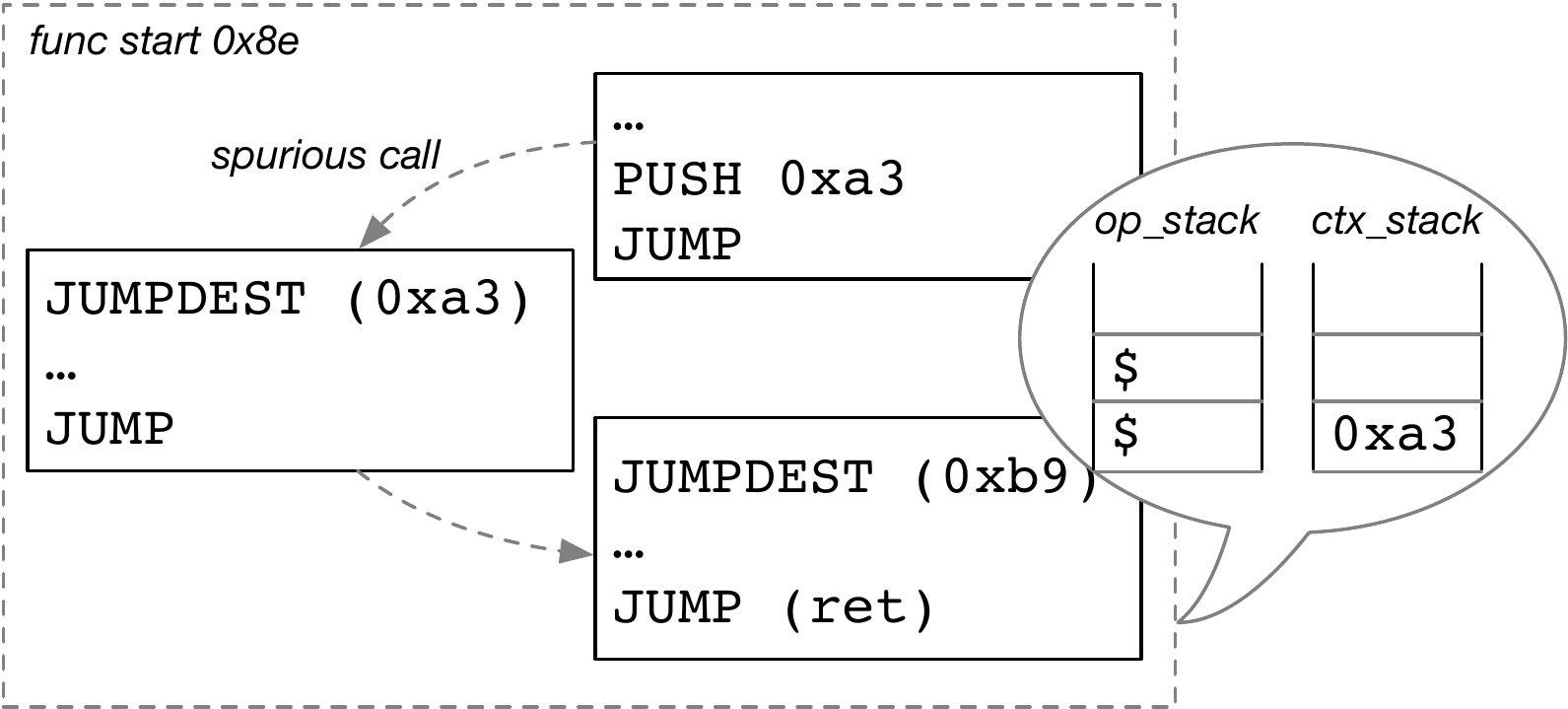}
	\caption{Invalid case: Recognizing the misjudged call-site.}
	\label{spurious call site}
\end{figure}
	
\begin{figure}[t]
	\centering
	\includegraphics[width=2.5in]{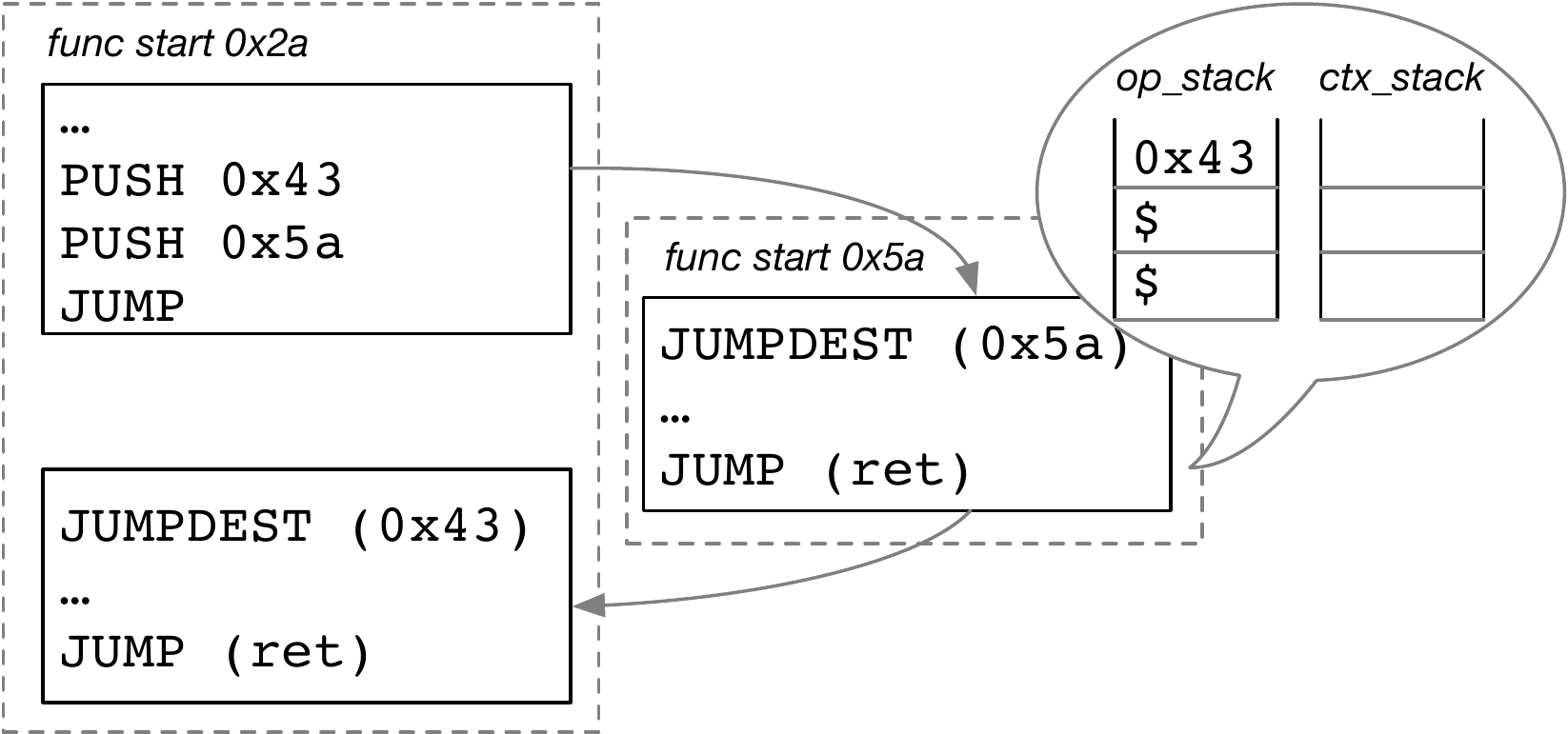}
	\caption{Invalid case: Recognizing a missing call-site.}
	\label{missing call site}
\end{figure}

\begin{enumerate}
	\item 
	The indirect jump uses the special symbol $\$$ in the operand stack as the return address but the $ctx$ is not empty. 
	This information shows that the return addresses used in the called function are not prepared for the caller. 
	This is a piece of strong evidence indicating that a control statement has been mistakenly recognized as an internal call (see Figure. \ref{spurious call site}). 
	As it is seen in Figure. \ref{spurious call site} the \textit{JUMP} targets to \textit{0xa3} are mistakenly recognized as an internal call resulting in the updated context stack. 
	The code entries at \textit{0xa3} are however not a function entry and the indirect jump reached from \textit{0xa3} is the exit of the caller (\textit{0x8e}), not the exit of function with entry \textit{0xa3}. This indirect jump uses the data which is not prepared for returning from function \textit{0xa3}. This is because $\$$ was prepared for the caller (\textit{0x8e}).
	
	\item 
	The indirect jump does not use the special symbol $\$$ in the stack as the address but the $ctx$ is empty. 
	If all internal call-sites are correctly identified in control flow traversal, the indirect jump must use $\$$ as the return address if $ctx$ is empty. 
	In this case, the contract uses the address prepared by the function as the return address. 
	This is an indicator that some internal call-sites were missed in the analysis.
	Therefore, we can lower the threshold $\rho$ to obtain the addresses with a lower probability as the candidate entries.
	This case is illustrated in Figure. \ref{missing call site}.
	The missing function entry at \textit{0x5a} results in misidentifying the related internal call-sites.
	After performing the intra-procedural jump into the missing function \textit{0x5a}, an indirect jump is performed, which is the exit of the missing function \textit{0x5a}.
	This indirect jump is mistakenly detected as the exit of the caller (\textit{0x2a}) and the indirect jump uses the constant value \textit{0x43} prepared by the caller. 
	It is evident that this indirect jump is not the exit of function with entry \textit{0x2a} and some function entries are missed in the analysis.
	
\end{enumerate} 

Using the above validation process for the function boundaries, we find some false positives and negatives for the function entries and internal calls. 
This information is further exploited in Algorithm .\ref{alg::static analysis} starting at the same entry to revise the identified boundaries.

\subsection{Addressing Challenges}
This section describes how neural-FEBI addresses the challenges raised in Section~\ref{secchal}.

First, although the EVM bytecode does not contain any symbolic information about the function, it still contains a few clues for the function entries identification of such stripped bytecode. 
The neural-FEBI utilized the well-known pattern to identify public function by analyzing the dispatcher and leveraged the FSI-LSTM-CRF to extract obscure clues of internal function entries from the EVM bytecode.

Note that the neural-FEBI does not need to identify internal function calls by establishing many complex patterns to cover all cases of internal function calls. 
In particular, it identifies the statement as a potential internal call if it jumps or falls through to a function entry predicted before.
The potential calls will be further filtered or added by the information of the operand and context stacks during control flow traversal to avoid false identification.
Furthermore, the entries of the function will be filtered when the function is never invoked.

Finally, the neural-FEBI deals with the overlapping function induced by instruction reshuffle, provided that there is a unique start address. 
Consider two functions that start at different addresses but contain the same bytes. 
During control flow traversal, the neural-FEBI will discover that both functions use the same bytes and attribute the bytes to both functions.

\section{Experiments}\label{experiments}
In this section, we evaluate and discuss the performance of our method. Specifically, we conduct experiments to address the following four research questions.

\begin{enumerate}[RQ.1:]
	\item How does the proposed neural-FEBI compare with existing approaches for the task of function entries identification in the stripped EVM bytecode?
	\item Does the proposed neural-FEBI perform better than the existing binary analysis tools for the task of function boundaries identification?
	\item What are the success rate and time consumption of neural-FEBI when analyzing smart contracts?
	\item How accurate are the intra-procedural CFGs and Call graphs built on top of the neural-FEBI?
\end{enumerate}

\begin{table*}[t]
\footnotesize
\renewcommand{\arraystretch}{1.3} 
\caption{Characteristics of the dataset used in experiments}
\label{neural:dataset}
\centering
\begin{tabular}{lllcll}
\hline
\multirow{1}*{} & \multicolumn{2}{l}{solc-0.4.25} &~& \multicolumn{2}{l}{solc-0.5.17} \\
\cmidrule(lr){2-3} \cmidrule(lr){5-6} 
& \bfseries optimized & \bfseries unoptimized &~& \bfseries optimized & \bfseries unoptimized \\
\hline

Number of binaries & 30,260 & 30,265 &~& 9,099 & 9,090 \\ 
\hline
Size (bytes) & 105,102,834 & 194,905,318 &~& 78,031,850 & 45,350,249 \\
\hline
Average Size (bytes) & 6,439.95 & 3,473.32 &~& 8,584.36 & 4984.09 \\
\hline
Number of functions & 660,561 & 666,112 &~& 279,772 & 288,771 \\
\hline
Average Number of functions & 21.8 & 22.0 &~& 30.74 & 31.76 \\
\hline

\end{tabular}
\end{table*}

\subsection{The Dataset and Ground Truth}

\begin{figure}[h]
	\centering
	\includegraphics[width=3in]{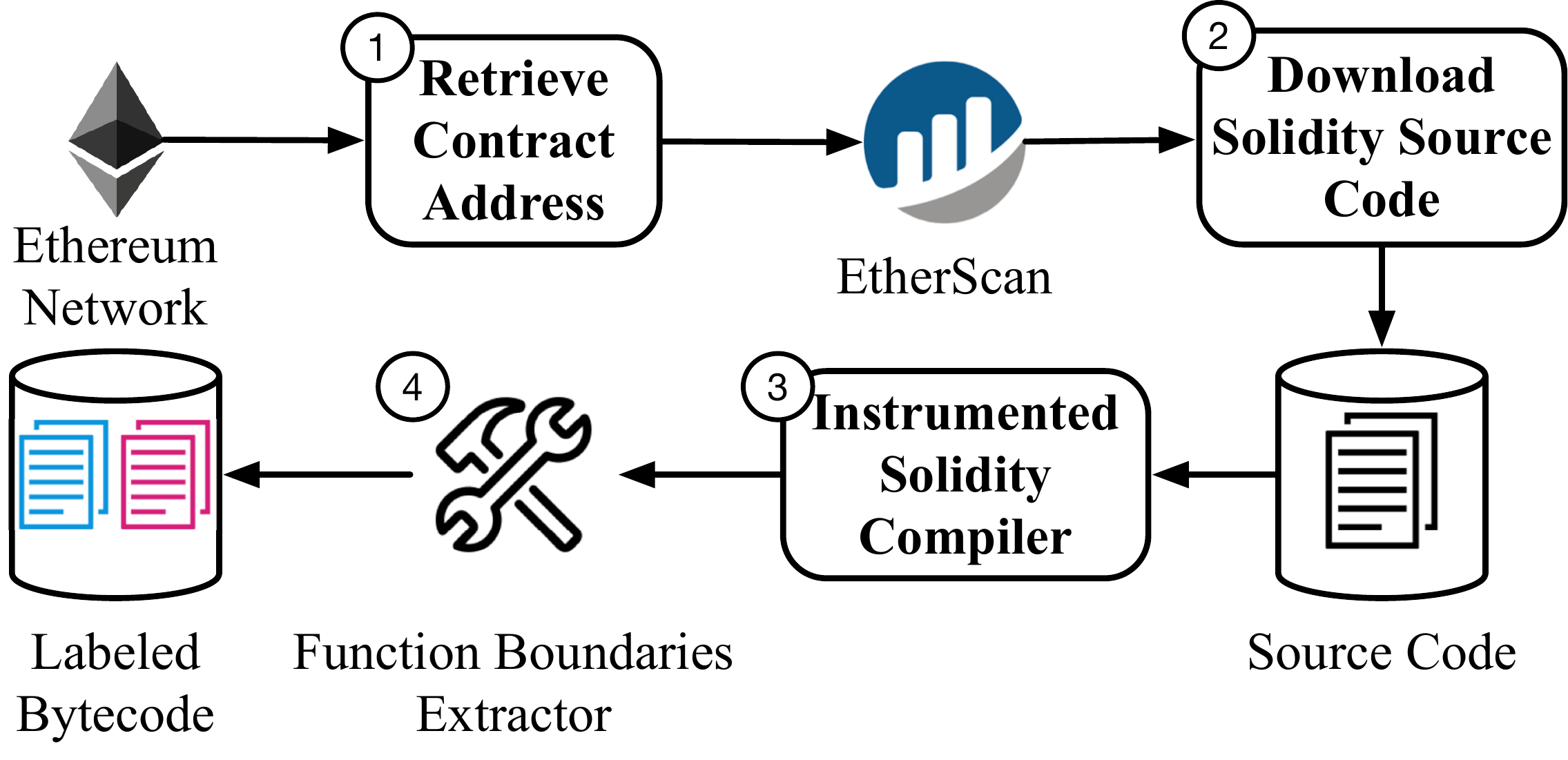}
	\caption{The generic workflow for smart contract acquisition and labeling}
	\label{gt_workflow}
\end{figure}

A public smart contract function identification dataset has not been available before. 
For this reason, we designed a tool with a generic workflow shown in Figure. \ref{gt_workflow} to gain the ground truth of function identification for stripped EVM bytecode. 
In step~(1), the addresses of the contracts are collected from the Ethereum mainnet. 
Step~(2) involves downloading the verified contract source code from EtherScan by its address. 
In step~(3), the source code of the smart contracts is compiled using a special instrumented Solidity compiler which can collect useful annotations about function boundaries. 
In the last step~(4), we developed a Python module to label function boundaries for stripped EVM bytecode by using the above collected annotations and output the ground truth. 
The developed toolchain and dataset are made publicly available online.

\subsubsection{Contract Acquisition}
For the dataset, we first utilize Geth API to collect the addresses of the contracts from Ethereum mainnet. 
We collected 5,801,258 contract addresses from the first 10 million Ethereum blockchain blocks. 
{However, we cannot} extract the ground truth of functions from these stripped binaries. 
To address this issue, we use the Ethereum API to collect the verified source codes of contracts which are then used to obtain the ground truth using the instrumented Solidity compiler. 
Since not all the source codes are available, we collected 1,189,567 source codes and 50,994 unique contracts.

\subsubsection{Ground Truth}
Studying the research questions requires establishing the ground truth for the function entries and their boundaries. 
After collecting the source code of the contracts, they are compiled using the instrumented compiler to obtain the ground truth. 

We instrumented the monitoring model in a Solidity compiler to collect the compilation optimization steps and the function boundaries before optimization.
The function boundaries are clear before optimization because the compiler translates the functions written in Solidity to the EVM bytecode one after another. 
With the information on optimization steps, we can track how the optimizer in the compiler modifies the function boundaries and constructs the boundaries for the corresponding optimized binary. 

Many compilers exist, so it is not practical to instrument all the existing compilers. 
We have compiled the 50,994 collected contracts by official compilers from \textit{solc-0.3.x} to \textit{solc-0.7.x}. 
The result shows that the \textit{solc-0.4.25} is the widely adopted version, which can successfully compile 59.2\% collected contracts. 
Besides the versions of \textit{solc-0.4.x}, the \textit{solc-0.5.17} is the most widely adopted one.
These two versions successfully compile about 76\% collected contracts.
For these reasons, we chose \textit{solc-0.4.25} and \textit{solc-0.5.17} to instrument.

To ensure the instrumented compiler does not have any impact on the experiment, we also compare the bytecode from the instrumented compiler with the original one.

\begin{table*}[t]
\small
\renewcommand{\arraystretch}{1.3} 
\caption{Comparison of different models for the internal function entries identification.}
\label{FSI}
\centering
\begin{tabular}{l lllc lllc lllc lll}
\hline
\multirow{3}*{Model} & \multicolumn{7}{l}{solc-0.4.25} &~& \multicolumn{7}{l}{solc-0.5.17} \\
\cmidrule{2-8} \cmidrule{10-16}
& \multicolumn{3}{l}{\bfseries optimized} &~& \multicolumn{3}{l}{\bfseries unoptimized} &~& \multicolumn{3}{l}{\bfseries optimized} &~& \multicolumn{3}{l}{\bfseries unoptimized} \\
\cmidrule{2-4} \cmidrule{6-8}  \cmidrule{10-12} \cmidrule{14-16} 
&  P &  R &  F1 &~&  P &  R &  F1 &~&  P &  R &  F1 &~&  P &  R &  F1 \\ 
\hline
CRF & 0.939 & 0.868 & 0.879 &~& 0.932 & 0.987 & 0.945 &~& 0.911 & 0.756 & 0.795 &~& 0.941 & 0.973 & 0.948 \\ 
\hline
biLSTM & 0.940 & 0.856 & 0.877 &~& 0.968 & 0.981 & 0.965 &~& 0.886 & 0.755 & 0.785 &~& 0.943 & 0.970 & 0.940 \\
\hline
biRNN & 0.943 & 0.891 & 0.902 &~& 0.980 & 0.988 & 0.980 &~& 0.906 & 0.761 & 0.808 &~& 0.962 & 0.964 & 0.958 \\
\hline
\hline
Elipmoc & 0.953 & 0.798 & 0.840 &~& 0.962 & 0.811 & 0.855 &~& 0.921 & 0.698 & 0.763 &~&  0.919 & 0.692 & 0.757 \\
\hline
Gigahorse & 0.935 & 0.702 & 0.759 &~& 0.944 & 0.647 & 0.714 &~& 0.815 & 0.562 & 0.629 &~&  0.820 & 0.496 & 0.569 \\
\hline
Vandal & 0.435 & 0.324 & 0.338 &~& 0.496 & 0.369 & 0.388 &~& 0.180 & 0.114 & 0.121 &~& 0.229 & 0.142 & 0.154 \\
\hline
\hline
\bfseries neural-FEBI & 0.977 & 0.932 & 0.946 &~& 0.998 & 0.998 & 0.997 &~& 0.900 & 0.886 & 0.883 &~& 0.997 & 0.997 & 0.996 \\
\hline
\end{tabular}
\end{table*}

\subsubsection{Dataset}
 The instrumented \textit{solc-0.4.25} successfully compiles 30,265 source codes in the unoptimized mode and 30,260 in the optimized mode. 
 The \textit{solc-0.5.17} also successfully compiles 9,090 in the unoptimized mode and 9,095 in the optimized mode. 
The union of such four datasets contains 38,996 contracts identified by their accounts corresponds to about 76\% of the collected contracts. 
The collected data is available on our project website, with Table. \ref{neural:dataset} summarizing the characteristics of the collected dataset.

\subsection{Comparison tools}
Among the existing Ethereum smart contracts decompilers, we select for comparison with neural-FEBI the ones that perform the task of function identification, including internal functions and private functions. This criterion led us to consider the following approaches:
\begin{itemize}
	\item Vandal: static analysis framework for Ethereum smart contract that decompiles bytecode to an intermediate representation that includes the private/public function boundaries.
	\item Gigahorse: decompilers that infer function boundaries heuristically for public and private functions from the bytecode. 
	\item Elipmoc: is essentially Gigahorse 2.0, which introduces more complete techniques, including transactional sensitivity and a fully context-sensitive private function reconstruction process. 
\end{itemize}
 

We discarded other tools such as \textit{EtherIR} \citep{EthIR}, \textit{Octopus} \footnote{https://github.com/FuzzingLabs/octopus}, \textit{EtherSolve} \citep{EtherSolve}, \textit{Panoramix} \footnote{https://github.com/palkeo/panoramix} and \textit{Evm bytecode decompiler} \footnote{https://github.com/MrLuit/evm} because they can not identiy the internal function. 
The \textit{ehtervm} \footnote{https://ethervm.io/decompile} can decompile contract bytecode into readable Solidity-like code, including the information of internal functions.
But it only emits the high-level code and we can not adopt this code in a common representation for further comparison.
We are not considering the JEB \footnote{https://www.pnfsoftware.com/jeb/evm} decompiler as they are closed-source tools that offer no way to perform extensive experimental evaluations in their free versions.

\subsection{Metrics}
\label{sec:metrics}
Given a set of function entries $S_p$ outputted by a method, we can define the true positive as $S_p \cap S_g$, where $S_g$ is the ground truth. 
Similarly, a false positive is defined as $S_p-S_g$, and a false negative is defined as $S_g-S_p$. 
We use the same metrics including $precision$, $recall$, $F1$ the previous works \citep{bao2014byteweight}. 
The $TP$ denotes the number of true positives, $FP$ denotes false positives, and $FN$ denotes false negatives.
 
\begin{alignat}{2}
	precision \; & = \; & & \frac{TP}{TP+FP} \\
	recall \; & = \; & & \frac{TP}{TP+FN} \\
	F1 & \; = \; & &\frac{2 \cdot precision \cdot recall}{precision+recall}
\end{alignat}

For the function boundaries identification task, we also define true positive, false positive, and false negative similar to the above{, but} all bytes of a function must completely match with the ground truth and the considered methods. 
All collected contracts are taken into account for evaluation, and we assign a zero score when tools fail to decompile.

\subsection{Performance of neural-FEBI}
In the experiments, we compare our results with the one generated by the state-of-the-art, marked as "Elipmoc", "Gigahorse", and "Vandal". 
The comparisons are based on the F1-score, precision, and recall metrics.
Because the neural-FEBI and the other learning-based methods need training, we divide the available data into training and testing sets. 
In our experiments, we randomly selected 50\% of collected contracts as training data and testing with the reminder, and we only compared the result of contracts in the testing sets with that of the existing tools. 
To demonstrate our effectiveness, we challenge our method against learning-based methods in the subtask of function entry identification. 
A 10-fold cross-validation was applied for each learning-based method to mitigate the non-determinism introduced by the random selection.

\begin{table*}[t]
\small
\renewcommand{\arraystretch}{1.3} 
\caption{Comparison of different learning models for the internal function entries identification by applying 10-fold validation (F1 scores).}
\label{FSI_cross_validation}
\centering
\begin{tabular}{l lc lc lc l}
\hline
\multirow{3}*{Model} & \multicolumn{2}{l}{solc-0.4.25} &~& \multicolumn{2}{l}{solc-0.5.17} \\
\cmidrule{2-4} \cmidrule{5-7}
& \multicolumn{1}{l}{\bfseries optimized} &~& \multicolumn{1}{l}{\bfseries unoptimized} &~& \multicolumn{1}{l}{\bfseries optimized} &~& \multicolumn{1}{l}{\bfseries unoptimized} \\
\hline
CRF & 0.881\;($\pm$0.011) &~& 0.947\;($\pm$0.004) &~& 0.798\;($\pm$0.014) &~& 0.952\;($\pm$0.017) \\ 
\hline
biLSTM & 0.904\;($\pm$0.015) &~& 0.985\;($\pm$0.003) &~& 0.811\;($\pm$0.032) &~& 0.968\;($\pm$0.009) \\ 
\hline
biRNN & 0.924\;($\pm$0.012) &~& 0.988\;($\pm$0.015) &~& 0.817\;($\pm$0.021) &~& 0.958\;($\pm$0.017) \\ 
\hline
\hline
neural-FEBI & 0.929\;($\pm$0.005) &~& 0.997\;($\pm$0.001) &~& 0.914\;($\pm$0.013) &~& 0.997\;($\pm$0.002) \\
\hline
\end{tabular}
\end{table*}

\subsubsection{RQ.1 - Function Entries Identification}
Table. \ref{FSI} summarizes our experimental results for the tasks of internal function identification by randomly selecting 50\% of contracts. 
Our proposed method consistently obtains F1-scores in the range of 88.3\%-99.7\% for different data sets. 
As expected, the model is best to run on binaries without optimization across different compiler versions. 
This is because such codes tend to include clear indications of the function entries. 
Running on the data compiled by \textit{solc-0.5.17} with optimization, the proposed framework only obtained 88.3\% in F1-score. 
The newer compiler may introduce advanced optimization techniques that generate more complicated code constructs. 
The larger code size in the \textit{solc-0.5.17} data set is another reason for this slight performance reduction.

Compared with the state-of-the-art, the neural-FEBI performance is better than Elipmoc, which achieves an F1-score in the range of 85.5\%-75.7\%.
During manual analysis, we discovered that Elipmoc or Gigahorse miss a few internal function entries because the call-sites targeted to them are not detected.
The Gigahorse had missed a few call-sites because they could not identify the call-site with complex arguments.
For example, we observed that the Vandal obtains less than 50\% F1-score in all data sets.
Although the Elipmoc can detect some complex arguments with potential call-return pairing, we found that the call-return filter implementation only retains the call-returns with high confidence and filters some correct function call-sites finally.

To the best of our knowledge, there is no other method based on learning to identify internal function entries in the EVM bytecode. 
Therefore we test the CRF, bidirectional LSTM, and bidirectional RNN on this task. 
The LSTM and RNN are used for similar tasks for x86 bytecode and achieve acceptable results. \citep{shin2015recognizing} 
Table. \ref{FSI} reports the precision, recall, and F1-score for different datasets on 50\% of random-selected contracts.
To mitigate the non-determinism introduced by the random selection, we also used standard ten-fold validation for these learning-based methods, dividing the element set into 10 sub-sets, applying 1 of the 10 on testing, and using the remaining 9 for training.
The results of this experiment are reported in Table. \ref{FSI_cross_validation}, revealing that neural-FEBI over-performs all of the above learning methods. 
\\
\\

\textit{Summary~(RQ.1): In the task of internal function entries identification, the proposed neural-FEBI archives comparable results to other approaches based on program analysis and supervised learning.}

 \begin{figure*}[t]
	\center
	\subfigure[0.4.25 optimized]{
		\label{unoptimized}
		\includegraphics[width=3in]{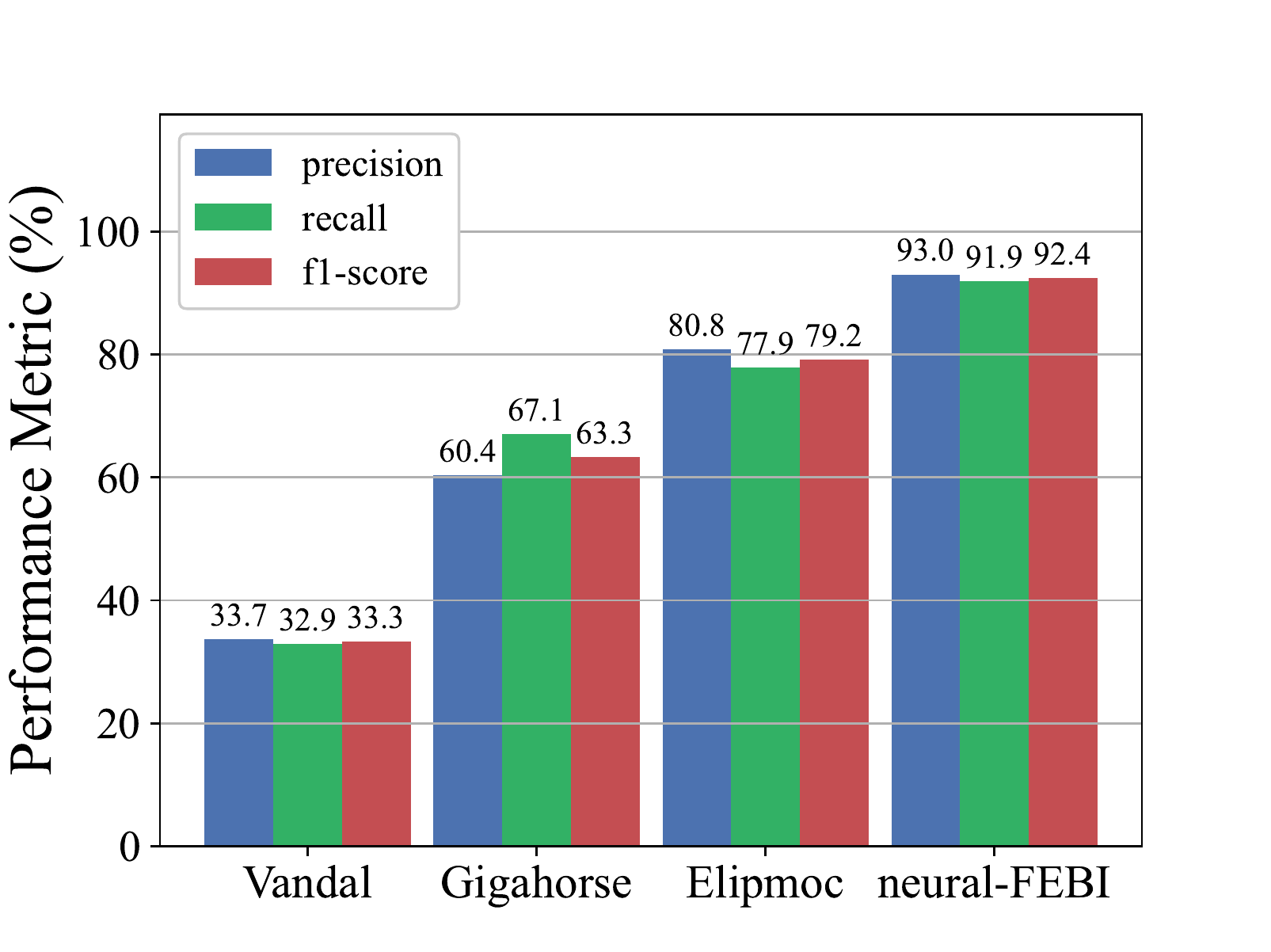}
	}
	\subfigure[0.4.25 unoptimized]{
		\label{optimized}
		\includegraphics[width=3in]{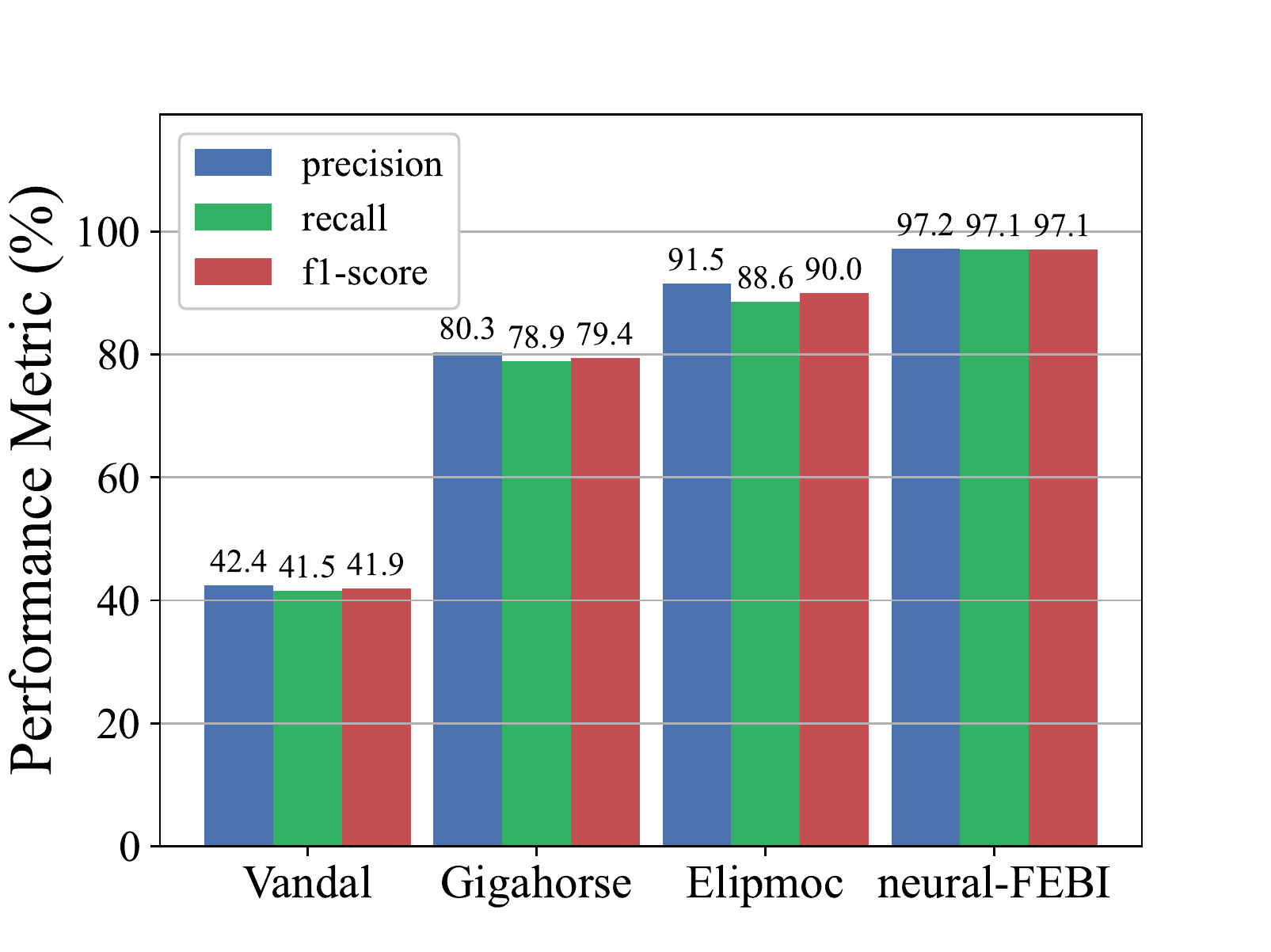}
	}
	\subfigure[0.5.17 optimized]{
		\label{unoptimized}
		\includegraphics[width=3in]{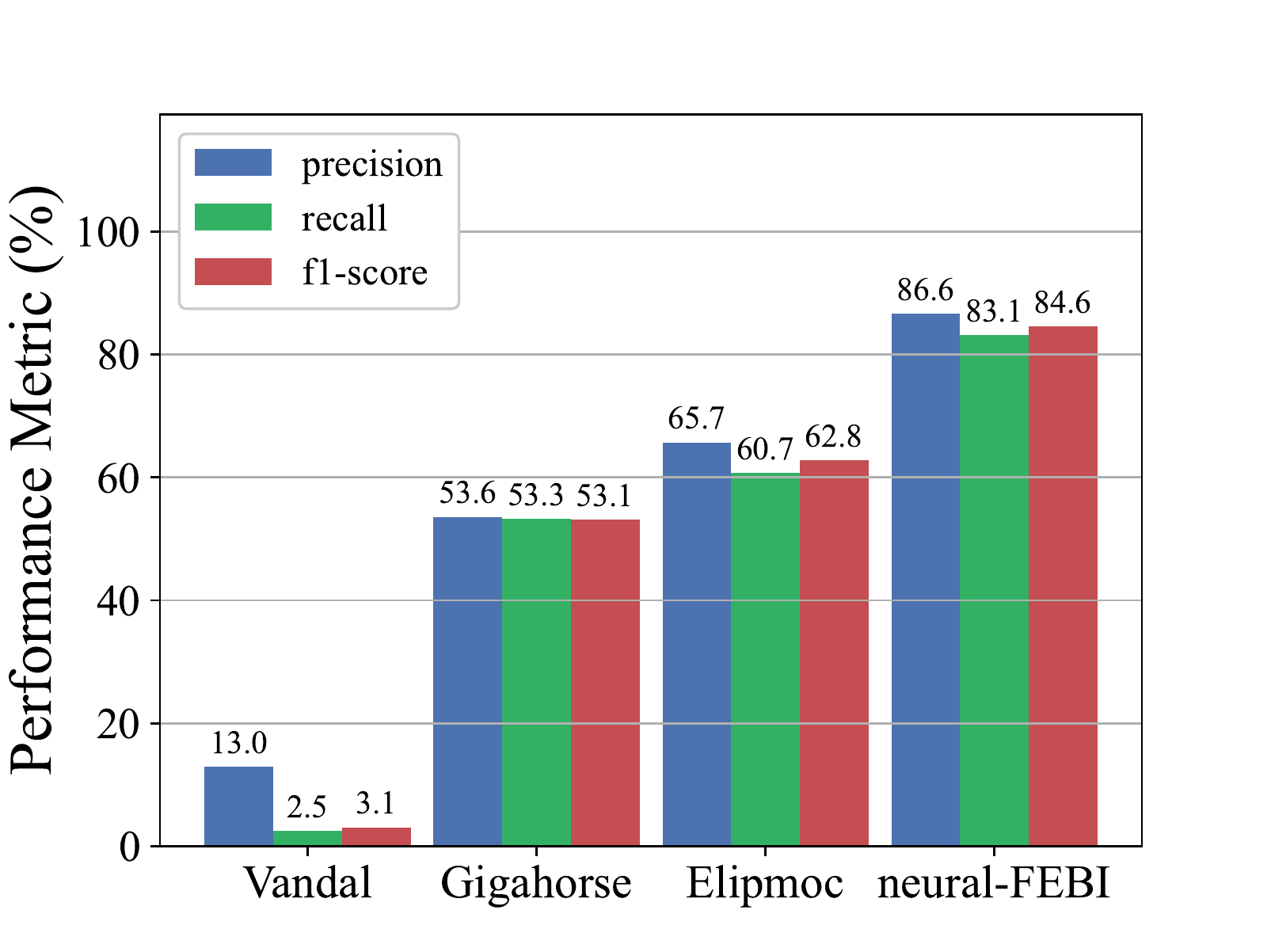}
	}
	\subfigure[0.5.17 unoptimized]{
		\label{optimized}
		\includegraphics[width=3in]{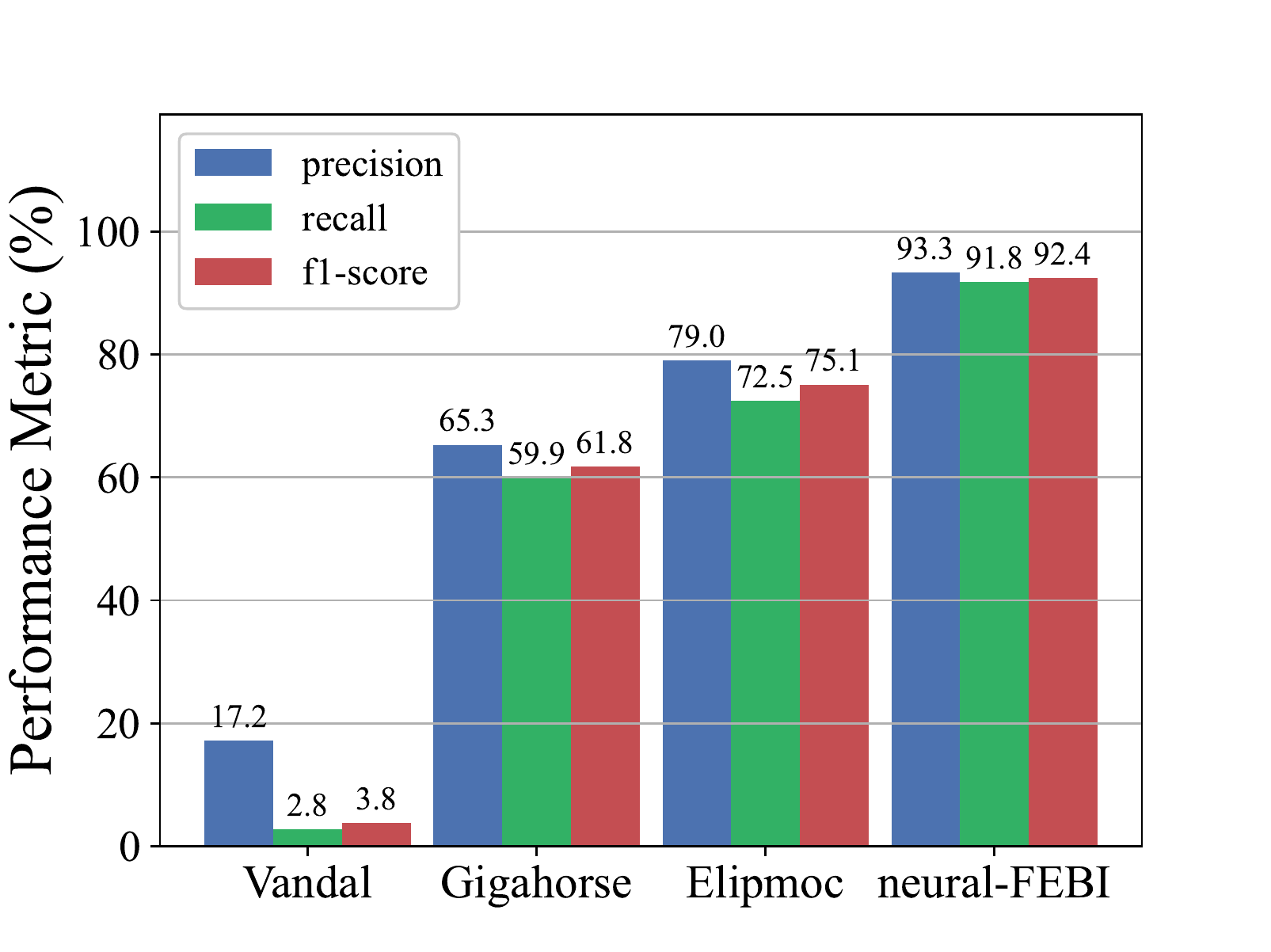}
	}
	\caption{Function boundaries identification: summary of our results and comparison with different tools.}
	\label{FBD}
\end{figure*}
 
\subsubsection{RQ.2 - Function Boundaries Identification}
The second experiment compares the performance of the proposed framework with the existing binary analysis tools for the function boundaries identification tasks on the EVM bytecode. 
Figure. \ref{FBD} presents a summary of the results, highlighting that the proposed neural-FEBI archive best results in  \textit{solc-0.4.25(unoptimized)} dataset and is comparable to other tools in a different dataset.

Additionally, the neural-FEBI archives more comparable results with other tools for optimized contracts than the unoptimized contracts.
In particular, for binaries compiled by \textit{solc-0.5.17}, neural-FEBI precision and recall are above 86.6\% and 83.1\% respectively, whereas for Elipmoc these values are 65.7\% and 60.7\%, respectively. 
By manually inspecting the function boundaries detected by Elipmoc, which have a lower f1-score, we verified that, in some cases, the reachable code of the called function is completely embedded in its caller. 
This is because the heuristics implemented in Elipmoc can not identify internal calls correctly in the previous stage. 
Meanwhile, the functions are omitted when they are not able to identify all of the internal calls that are aimed at them, and the boundaries of these functions have an omission.

Interestingly, it is also seen that Vandal is much worse than the others as it only identifies less than 10\% public function entries in the binaries compiled by \textit{solc-0.5.17}, whereas the other two tools can identify almost 100\% of public function entries. 
\\
\\

\textit{Summary~(RQ.2): The function boundaries identified by neural-FEBI are more accurate than previous tools, especially in the optimized contracts.}

\begin{figure*}[t]
	\center
	\includegraphics[width=6.4in]{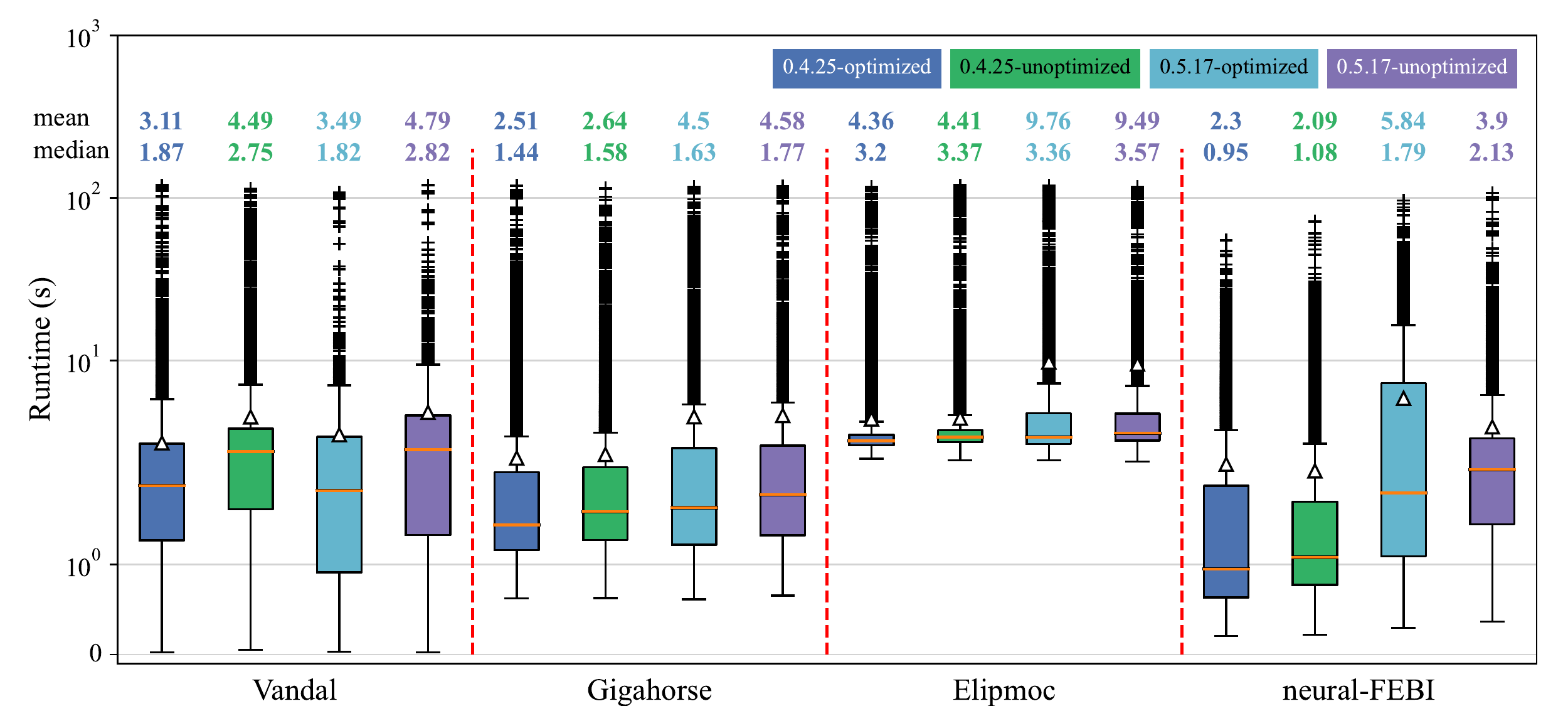}
	\caption{Wall-clock time taken to function identification on 50\% collected contracts (excluding contracts that were not successfully detected due to timeouts or fatal errors)}
	\label{performances_fig}
\end{figure*}

\subsubsection{RQ.3 - Success Rate and Efficiency}
In this section, we describe some experimental results in order to gain insight of the effectiveness of the proposed neural-FEBI.

\textbf{Timeouts/fatal errors.}
As shown in Table. \ref{performances_table}, both Gigahorse and Vandal failed to decompile a significant portion of the contracts subject to the 120s cutoff, especially for the contracts compiled by \textit{solc-0.5.17}. 
In contrast, neural-FEBI can successfully analyze almost all the contracts. 
The neural-FEBI timed out in less than 1\% of the contracts across different datasets. 
Although Gigahorse obtains satisfactory results with an earlier version of the compiler, it fails to analyze about 10\% of contracts based on the newer compiler. 
Furthermore, the enhanced version of Gigahorse, Elipmoc, archives acceptable success rates across different datasets, which can successfully analyze more than 96\% of contracts.
It is a strong indicator that the function inference in the proposed neural-FEBI is much more reliable than the others. 

\begin{table}[h]
\small
 \renewcommand{\arraystretch}{1.3} \caption{Successfully analyzed contracts for testing on different datasets.} 
\label{performances_table}
\centering
\begin{tabular}{lllcll}
\hline
\multirow{2}*{Approach} & \multicolumn{2}{l}{solc-0.4.25} &~& \multicolumn{2}{l}{solc-0.5.17} \\
\cmidrule(lr){2-3} \cmidrule(lr){5-6} 
& \bfseries opt & \bfseries unopt &~& \bfseries opt & \bfseries unopt \\
\hline
Elipmoc &  99.5\%  & 99.4\%  &~& 96.9\%   & 97.8\%  \\
\hline
Gigahorse & 98.5\%  & 98.6\%  &~& 90.7\%  & 91.5\%  \\
\hline
Vandal &  64.0\%  & 67.8\%  &~& 36.7\%   & 41.5\%  \\
\hline
\bfseries neural-FEBI & 99.8\%  & 99.9\%  &~& 99.3\%  & 99.0\%  \\
\hline
\end{tabular}
\end{table}

\textbf{Time consuming.}
The distribution of runtime performance is presented in Figure. \ref{performances_fig}, revealing that neural-FEBI is very fast for all contracts. 
For example, each contract requires, on average, 2.3s for function boundary identification in the \textit{solc-0.4.25(optimized)} dataset. The only exception is the \textit{solc-0.5.17(optimized)} dataset. 
The reason is the neural-FEBI attempts to lower the threshold to catch the underlying information of the missing call-sites due to the complicated structure of the contracts. 
Each time lowing the threshold, the control flow analysis starts from candidate function entries and needs to be performed again, which is the most time-consuming in the proposed neural-FEBI. 
Moreover, the results are shown in Figure. \ref{performances_fig} are only including the successful ones and the neural-FEBI successfully identify almost more 10\% contracts than Gigahorse. 
For these 10\% contracts, most are large-size contracts and Gigahorse failed because more diverse contracts that warrant analysis efforts are greater.
\\
\\

\textit{Summsary~(RQ.3): Neural FEBI is able to analyze 99\% of collected contracts in our experiment, and it takes less time to process them.}

\subsubsection{RQ.4 - Call Graphs and Intra-procedural CFGs}

We use two applications to demonstrate the usefulness of neural-FEBI for improving the result of reverse engineering, including call graphs and intra-procedural CFGs construction. 
Due to the accurate function boundaries provided by neural-FEBI, the reliability of these applications can be boosted. 

Building a CFG from EVM bytecode describes the basic structure of smart contracts. 
Most previous tools, such as EhterSolve, EthIR and Octopus, only split the binaries into basic blocks and constructed the global CFG without information about functions.  
However, these global CFGs are difficult to comprehend because all functions are fused into a huge and messy direct graph.
Assigning source function semantics to such global CFG is necessary for the programmer to understand the program in terms with which they are familiar, provide proper labeling when the programmer has the source code, and provide concrete targets for program instrumentation and modification \citep{meng2016binary}.
The function boundary splits the whole global CFG into an amount of intra-procedural CFGs and uses a call graph to show the invocation relationship between them.
These graphs are vital for smart contract context and path profiling. 
Specifically, the smart contract intrusion detection system, ContractGuard, detects abnormal behavior by profiling context-tag acyclic paths \citep{wang2019contractguard}. 
Without the accurate call graph and intra-procedural CFGs, the profiling algorithms leveraged by the ContractGuard may provide improper results for the following embedding instrumentation.

We extended the neural-FEBI to construct such graphs,  apply it to all collected contracts across different datasets, and compare the results with the ground truth from the instrumented Solidity compiler. 
Each intra-procedural CFG contains a list of nodes, identified by the first offset (entry) of the basic block and a list of edges, identified by a pair of offsets.
For each smart contract, we compare the acyclic paths, a sequence of offsets representing the entry of basic block, among the intra-procedural CFGs generated by different tools.
Informally, the intra-procedural acyclic path is a path on the acyclic version of CFG, which is obtained by replacing every backedge \textit{w}$\rightarrow$\textit{v} with two surrogate edges \textit{entry}$\rightarrow$\textit{v} and \textit{w}$\rightarrow$\textit{exit}. 
Finally, we used DFS to find all paths from \textit{entry} to \textit{exit} in acyclic intra-procedural CFGs for comparison.

Table \ref{cfg_table} gives the F1-score of the acyclic paths prescribed by the intra-procedural CFGs constructed by the Elipmoc, Gigahorse and our approach. 
The results confirm that our approach constructs significantly more accurate intra-procedural CFGs than those by Elipmoc and Gigahorse.

\begin{table}
\small
 \renewcommand{\arraystretch}{1.3} \caption{ Intra-procedural CFGs: summary of our results and comparison with state-of-the-arts.} 
\label{cfg_table}
\centering
\begin{tabular}{lllcll}
\hline
\multirow{2}*{Approach} & \multicolumn{2}{l}{solc-0.4.25} &~& \multicolumn{2}{l}{solc-0.5.17} \\
\cmidrule(lr){2-3} \cmidrule(lr){5-6} 
& \bfseries opt & \bfseries unopt &~& \bfseries opt & \bfseries unopt \\
\hline
Elipmoc & 74.7\%  & 81.7\%  &~& 57.8\%  & 68.5\%  \\
\hline
Gigahorse & 71.6\%  & 75.3\%  &~& 51.6\%  & 54.4\%  \\
\hline
\bfseries neural-FEBI & 86.5\%  & 98.5\%  &~& 78.4\%  & 93.3\%  \\
\hline
\end{tabular}
\end{table}

\begin{table}
\small
 \renewcommand{\arraystretch}{1.3} \caption{Call graph: summary of our results and comparison with state-of-the-arts.} 
\label{call_graph_table}
\centering
\begin{tabular}{lllcll}
\hline
\multirow{2}*{Approach} & \multicolumn{2}{l}{solc-0.4.25} &~& \multicolumn{2}{l}{solc-0.5.17} \\
\cmidrule(lr){2-3} \cmidrule(lr){5-6} 
& \bfseries opt & \bfseries unopt &~& \bfseries opt & \bfseries unopt \\
\hline
Elipmoc & 94.4\%  & 95.1\%  &~& 84.5\%  & 85.3\%  \\
\hline
Gigahorse & 91.7\%  & 91.7\%  &~& 76.8\%  & 76.7\%  \\
\hline
\bfseries neural-FEBI & 98.4\%  & 99.5\%  &~& 91.8\%  & 92.8\%  \\
\hline
\end{tabular}
\end{table}

The representation of the call graph contains a list of nodes identified by function entries and the call-sites as edges. 
We extend the call graph with a unified enter identified by the dispatcher and connect it to all entries of the public function.
A path set from the dispatcher to each node was exploited for the comparison of call graphs.
 
Due to recursion, the call graph may contain cycles. 
In the same way as a backedge is replaced by a subrogate edge in a CFG, each recursive call-site is replaced by a subrogate call-site from the dispatcher to the function that is called recursively. 
Table \ref{call_graph_table} presents the experimental results based on the F1-score, demonstrating that the call graph constructed by our technique significantly outperforms the competitor methods.

\textit{Summary~(RQ.4): The neural-FEBI is able to build more precise intra-procedural CFGs and call graphs with the aid of function boundaries than other previous tools have shown.}

\begin{figure}[h]
	\centering
	\includegraphics[width=3in]{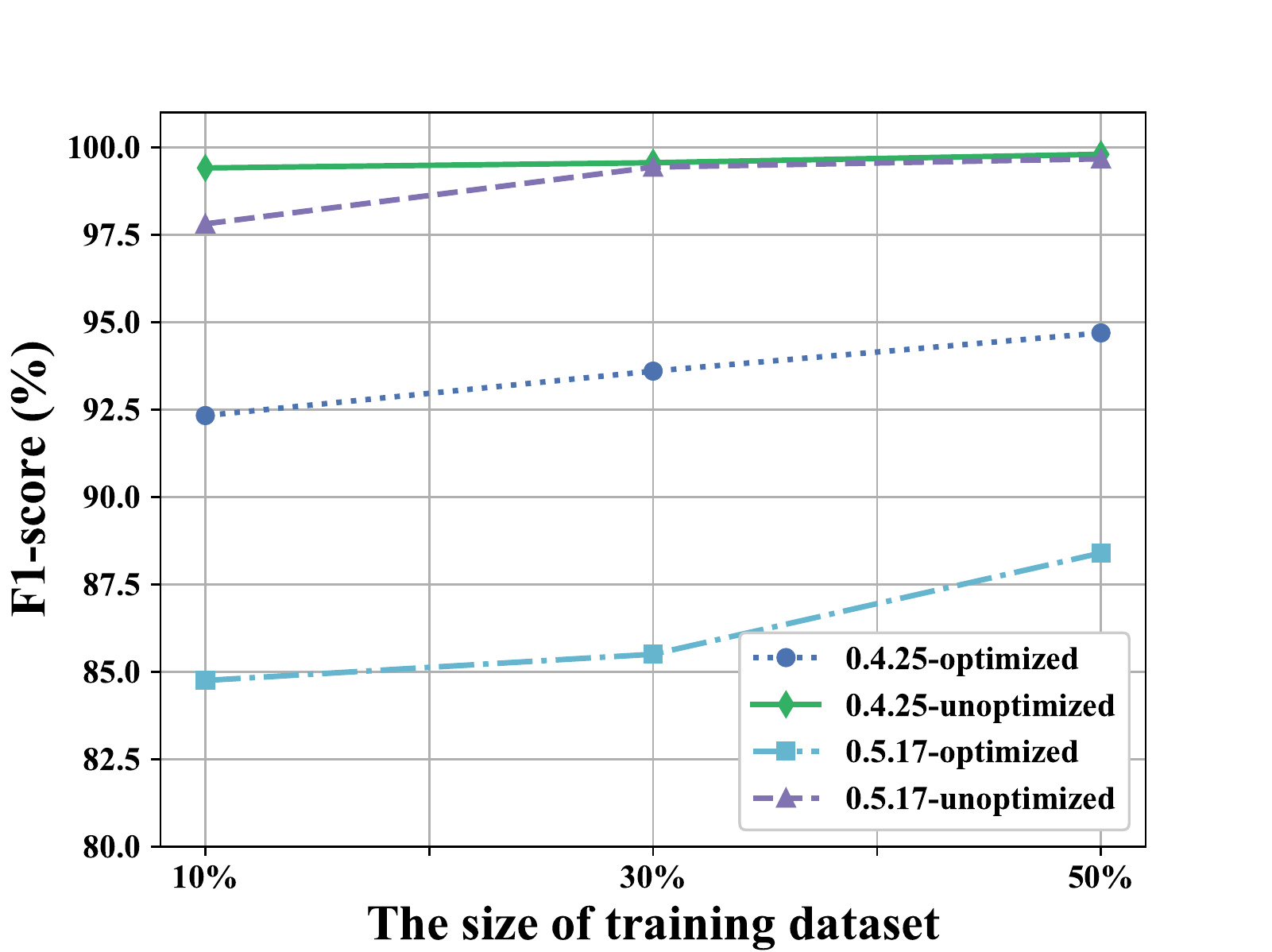}
	\caption{Results with when trained on 10\%, 30\%, 50\% of the data (F1-score).}
	\label{fsi_diff_size}
\end{figure}

\subsection{Threats to validity}
Here we discuss the following several issues arising in the implementation and experiments with the proposed neural-FEBI framework.
\begin{enumerate}[1.]
	\item
	An important limitation of the proposed neural-FEBI is that the training data sets are required, while other comparable tools are not. 
	As such, there is vital to simulate common applications of binary analysis where only a small amount of training data is available. 
	We evaluate the performance of neural-FEBI for the task of function entries identification by training 10\% and 30\% of the binaries in the dataset and testing on the rest. 
	Interestingly, the performance is only reduced by 3.6\% at most, as shown in Figure. \ref{fsi_diff_size}.
	
	\item 
	It's worth noting that the body entries of the public function are the same as the internal function entries. 
	However, the body entries in the latter can be entered from the dispatcher. 
	Hence, the recognition of the body entries has a strong incentive as it discovers the internal function entries. 
	Since these two tasks are strongly related, in the training process, we combine the task of body entry identification and internal entry identification. 
	This fits with the setting of multi-task learning and transfer learning. 
	To mediate these two tasks, during the training process, we also flag the body entries together with the internal function start to the same label. 
	This ensures that our framework can provide the related knowledge of body entries to the task of internal function entry identification.
	 
	\item 
	In the EVM bytecode, the \textit{JUMPDEST} is the only valid instruction for function entries. It is also seen that some functions are not started with \textit{JUMPDEST}. In cases where the functions are only called once and their entries can be reached by fall-through, the entries denoted by \textit{JUMPDEST} may be discarded by \textit{Peephole} optimization. More specifically, the above functions are in-lined in the caller function as these functions are not called by other functions anymore. In other words, their boundaries are fusing into the caller function. In this paper, we excluded in-lined functions.
	
	\item 
	In some contracts, the number of functions in the Solidity program is not equal to that in the corresponding EVM bytecode. 
	This is because the compiler may remove the duplicated functions and import simple built-in functions. 
	In this paper, such functions are considered as they have the same features as the functions declared in the source code. 
	The duplicated functions are however ignored since they are removed from the EVM bytecode.
	
	\item 
	As with most learning-based approaches, the neural-FEBI will fail to correctly identify the entries of function if similar patterns from the training data do not exhibit themselves in the test data. 
	New compilers may introduce more advanced techniques to reduce the size of contracts, which increases the complexity of function identification, and the performance of neural-FEBI may be slightly reduced.
	Besides, smart contracts can be interleaved with inline assembly statements for more fine-grained. 
	These statements are extremely low-level and easily re-arranging functional-style opcodes, making the neural-FEBI unable to identify the functions.

\end{enumerate}

\section{Related Works}\label{relatedworks}

\subsection{Decompilers for Smart Contracts}
The popularity of the Ethereum platform and the fact that the most deployed smart contracts do not have readily linkable source code available resulted in the emergence of many EVM bytecode decompilation tools.
They can be classified according to the underlying techniques.
Some of them rely on symbolic execution to decompile EVM bytecode. 
For example, tools including EthIR \citep{EthIR}, EtherSolve \citep{EtherSolve} and Mythril\citep{mueller2018smashing} with the aid of a symbolic execution engine to explore the traces of complied smart contracts.
SigRec \citep{SigRec} is proposed by leveraging type-aware symbolic execution to detect known patterns used by Solidity compiler to recover function signatures automatically. 
Other works rely on program analysis. 
For example, Eray \citep{zhou2018erays} uses program analysis to produce high-level pseudocode suitable for manual analysis, the Vandal \citep{brent2018vandal} is an open-source framework written by Python which decompiles EVM bytecode into IRs.
Recently, existing decompilers have used declarative theorem proving. 
For example, Gigahorse \citep{grech2019gigahorse} decompiles smart contracts from EVM bytecode into a high-level 3-address code representation, which implicit data- and control-flow dependencies of the EVM bytecode.
Elipmoc \citep{Elipmoc} is an evolution of Gigahorse by employing several high-precision heuristics and making them scalable.
There are also some tools proposed in the industry to recover solidity-like programs from the bytecode of smart contracts, such as Octopus, Panoramix, ethervm, EVM bytecode decompiler and JEB. 

Most of the tools mentioned above can infer public functions from EVM bytecode. In contrast to private functions, public functions are relatively easy to identify because they have well-known entry points. 
To our best knowledge, the Vandal, Gigahorse, Elipmoc and ethervm are the only others to try to identify private functions.

\subsection{Functions Identification}
A plethora of similarly sophisticated tools for other conventional bytecode have been developed over the years, both in academia \citep{brumley2011bap, meng2016binary, rosenblum2008learning, bao2014byteweight, shin2015recognizing, jima} and industry (i.e., IDA Pro, Jakstab and OllyDbg). 
For instance, Jima\citep{jima} proposed identifying functions for 32 and 64-bit Intel binaries by using static analysis and limited behavioral analysis. 
Some recent works have introduced supervised learning techniques and neural networks to identify the functions automatically.\citep{rosenblum2008learning, bao2014byteweight, shin2015recognizing}.
For instance, \citep{bao2014byteweight} proposed BYTEWEIGHT, a machine-learning-based approach for the x86 function entries identification task. 
\citep{shin2015recognizing} also shows that the recurrent neural network \citep{elman1990finding} can identify functions in binaries with higher accuracy and efficiency. 

It is important to realize that decompilation over conventional bytecode is technically an entirely different problem from EVM decompilation.
Conventional bytecode (i.e., x86) contains large amounts of high-level information (calls and jumps to known labels). 
In some ways even easier to identify functions once a reliable disassembly is produced.

\subsection{Security for smart contract}
In the literature, a number of tools have been proposed to verify smart contracts due to the security issues inherent in the high-risk paradigm of smart contracts.
For example, Oynete \citep{luu2016making}, Teether \citep{krupp2018teether}, Manticore \citep{maticore}, and SAFEVM \citep{safevm} introduce symbolic execution to detect if known vulnerabilities can be triggered on a control flow path. 
Security \citep{securify} also uses symbolic analysis of the contract’s dependency graph to extract precise semantic information from the code. 
TokenScope \citep{tokenscope} is also an automatic detector of inconsistent behaviours of the cryptocurrency tokens in Ethereum. 

There are also recent research works on binary rewriting techniques for improving the security of contracts. 
For example, EVMPatch \citep{rodler2021evmpatch} is proposed to instantly and automatically patch faulty contracts. 
ContractGuard \citep{wang2019contractguard} introduces a profiling technique on acyclic paths to protect contracts against unknown attacks that induce abnormal control flow after deployment. 
Some tools \citep{jiang2018contractfuzzer, EVMFuzzer, harvey} generate fuzzing inputs based on the ABI specifications of the smart contracts.

\section{Conclusion}\label{conclusion}
From the administrators’ perspective, Ethereum smart contracts are vulnerable to attacks after deployment. 
The existing analyses of the stripped EVM bytecode are often based on function entries and boundaries. 
The use of stripped EVM bytecode also introduces new challenges. 
In this paper, we propose Neural-FEBI to identify functions in the stripped EVM bytecode. 
Our evaluation of the real-world Ethereum contracts across different compiler versions and optimization levels confirms the efficiency and accuracy of the proposed method in identifying the function entries and boundaries for the internal functions. 
Our experimental results also indicate significantly higher performance than state-of-the-art decompilers, including Elipmoc, Gigahorse and Vandal. 
Moreover, we demonstrate that the recovered function boundaries can be used to construct more accurate intra-procedural CFGs and call graphs.

\section{Acknowledgement}\label{acknowledgement}
This work is supported by 
National Natural Science Foundation of China (No. 62006083), 
Science and Technology Projects in Guangzhou (No. 202102020654),
Bejing Municipal Science \& Technology Commission, Administrative Commission of Zhongguancun Science Park, 2022, Fintech Innovation Support Project (No. Z221100001222004)
Hong Kong RGC/GRF grant 16205821 and RGC/RIF R5034-18,
National Social Science Fund of China (No. 19ZDA041),
Key R\&D Program for Guangdong (No. 2019B010137003),   
and Industry innovation fund of Chinese University (No. 2020ITA09006).


\bibliographystyle{cas-model2-names}

\bibliography{./manuscript.bbl}

\end{document}